# Genarris: Random Generation of Molecular Crystal Structures and Fast Screening with a Harris Approximation

Xiayue Li[1,2], Farren S. Curtis[3], Timothy Rose,[1] Christoph Schober,[4] Alvaro
aVazquez-Mayagoitia[5], Karsten Reuter[4], Harald Oberhofer[4], Noa Marom[1,3,6,a)]

[1]*Department of Materials Science and Engineering, Carnegie Mellon University, Pittsburgh, PA 15213, USA*

[2]*Google Inc., Mountain View, CA 94030, USA.*

[3]*Department of Physics, Carnegie Mellon University, Pittsburgh, PA 15213, USA*

[4]*Chair for Theoretical Chemistry and Catalysis Research Center, Technische Universiät München, Lichtenbergstr. 4, D-85747 Garching, Germany*

[5]*Argonne Leadership Computing Facility, Argonne National Lab, Lemont, IL 60439, USA.*

[6]*Department of Chemistry, Carnegie Mellon University, Pittsburgh, PA 15213, USA*

We present Genarris, a Python package that performs configuration space screening for molecular crystals of rigid molecules by random sampling with physical constraints. For fast energy evaluations Genarris employs a Harris approximation, whereby the total density of a molecular crystal is constructed via superposition of single molecule densities. Dispersion-inclusive density functional theory (DFT) is then used for the Harris density without performing a self-consistency cycle. Genarris uses machine learning for clustering, based on a relative coordinate descriptor (RCD) developed specifically for molecular crystals, which is shown to be robust in identifying packing motif similarity. In addition to random structure generation, Genarris offers three workflows based on different sequences of successive clustering and selection steps: the "Rigorous" workflow is an exhaustive exploration of the potential energy landscape, the "Energy" workflow produces a set of low energy structures, and the "Diverse" workflow produces a maximally diverse set of structures. The latter is recommended for generating initial populations for genetic algorithms. Here, the implementation of Genarris is reported and its application is demonstrated for three test cases.

## I. INTRODUCTION

Understanding the solid-state behavior of molecules may inform the design of crystal forms with desired properties for target applications. Traditionally a prime interest of the pharmaceutical industry, molecular crystals also have applications in diverse areas such as solar cells,[1] organic light emitting diodes (OLEDs),[2] and porous materials for gas storage and catalysis.[3,4] Molecular crystals often display polymorphism, the ability of a molecule to crystallize in

---

[a)] Author to whom correspondence should be addressed. Electronic mail: nmarom@andrew.cmu.edu.

more than one structure.[5–7] Polymorphs of pharmaceuticals may exhibit significantly different physical and chemical properties such as stability, solubility, and processability.[5,8,9] For organic semiconductors, different polymorphs may display different band structures, optoelectronic properties, and electron–phonon couplings.[10–15]

Crystal structure prediction (CSP) is a grand challenge for the computational condensed matter community because it requires screening a large number of candidate crystal structures with high accuracy.[16–20] Sampling the configuration space for a given molecule is enormously complex, as one must consider a range of all possible space groups, lattice parameters, values of Z (the number of asymmetric units related by symmetry in the unit cell) and Z' (the number of molecules in the asymmetric unit), molecular orientations, and conformations. Furthermore, weak van der Waals interactions in molecular crystals lead to many local minima that are extremely close in energy, requiring energy resolution of a few meV for accurate ranking of polymorphs.[6,21–25] The progress of the field has been periodically assessed by CSP blind tests, organized by the Cambridge Crystallographic Data Centre (CCDC).[26–31] Over the course of six blind tests, spanning nearly two decades, several best practices have emerged for the generation and ranking of molecular crystal structures.

For ranking of putative structures, hierarchical screening approaches are often used, where successive steps employ increasingly accurate energy methods for smaller subsets of structures. Generic force fields have consistently been demonstrated to produce poor results in crystal structure prediction.[29–31] Tailor-made, system-specific force fields parameterized based on *ab initio* calculations have proven more reliable. Dispersion-inclusive density functional theory (DFT) has become the de facto standard for the final ranking of structures.[31] The many-body dispersion (MBD) method, in particular when combined with hybrid DFT functionals, has been shown to be highly accurate.[23,31–35] Fully *ab initio* calculations, however, are too computationally expensive for fast initial screening of a large number of structures. Parameterization or machine learning of tailor-made system specific interatomic potentials may also require a significant number of first principles calculations.

The Harris approximation (HA)[36,37] is a transferable first principles approach with a moderate computational cost that offers a compromise between the efficiency of empirical force fields and the accuracy of *ab initio* DFT calculations. Contrary to force fields or semi-empirical methods, the HA is entirely parameter free and can thus also readily be applied to entirely novel systems. Within the HA, the total density of a system is constructed by superposition of self-consistent fragment densities. The DFT total energy is then calculated for the Harris density without performing a self-consistent cycle.[36–38] The HA has been shown to perform well for weakly interacting molecular dimers, where there is no electron density overlap and no significant polarization.[38] To the best of our knowledge, here the HA is used for molecular crystal configuration space screening for the first time. In this case, the fragments are the constituent molecules of the crystal.

Random sampling of the configuration space is widely used in the structure generation process.[29,30,39–41] While some of the early pioneers of CSP used purely random or grid searches,[39,40,42] quasi-random sampling using low-discrepancy Sobol sequences provides a more uniform coverage.[31,43–45] Random sampling is often constrained by symmetry, stoichiometry, knowledge of the chemical system, and experimental data.[20,31] Random sampling frequently precedes or is incorporated into more advanced search algorithms,[31] such as genetic algorithms (GAs),[46–48] swarm algorithms, and Bayesian optimization. Several CSP methods rely on random structural modifications, including simulated annealing,[49,50] parallel tempering,[51] and basin hopping.[52–54] Random sampling is often combined with clustering methods to monitor the sampling convergence, as in the conformation family Monte Carlo method[55] and other quasi-random sampling techniques.[42,56]

Recently, data driven approaches, such as machine learning (ML) algorithms have been increasingly employed in computational chemistry and materials science in conjunction with first principles simulations,[57,58] in various capacities, including predicting a material's structure[59–61] and properties,[62–74] generating interatomic potentials[75–81] and DFT functionals,[82] improved sampling,[83–85] revealing structure-property correlations,[86–88] and finding predictive descriptors.[89–92] We expect ML to be featured heavily in the next CSP blind test. In particular, best practices for configuration space screening may benefit from using ML to perform (dis)similarity analysis while effectively capturing the similarity and diversity of crystal packing motifs. To this end, one widely used descriptor is the radial distribution function (RDF).[40,56,93] Other descriptors are based on a series of interatomic distances representing specific close intermolecular contacts.[41,56] Both of these descriptors are based on atomic positions. To capture the packing motifs of molecular crystals, we introduce a new relative coordinate descriptor (RCD), based on the relative positions and orientations of neighboring molecules.

Genarris is a Python package that currently performs configuration space screening for crystals of rigid molecules. It is available for download from www.noamarom.com under a BSD3 license. The purpose of Genarris is not necessarily to seek the ultimate convergence of the search (i.e. the global minimum structure), but rather to provide a computationally efficient way of generating a diverse set of reasonable structures that span the potential energy landscape. Genarris was originally developed in order to produce an initial population for the GAtor genetic algorithm package.[48] However, it may be applied more broadly to generate structure sets for any other search algorithm, for fitting system specific interatomic potentials, or for training machine learning algorithms. Genarris generates random structures with physical constraints imposed on symmetry, unit cell parameters, and intermolecular close contacts. The HA is then used for fast energy evaluations. Once a large "raw" pool of random structures has been generated, Genarris offers three standard workflows for further refinement. The "Energy" workflow selects for low energy structures. The "Diverse" workflow favors structural diversity over energetic stability. The "Rigorous" workflow involves hierarchical screening of structures and is essentially a CSP method in and of itself. All workflows incorporate ML using RCD-based clustering. The user may choose the most appropriate workflow, depending on their needs and computational resources. In the following, we report the implementation of Genarris, validate the reliability of the HA and the effectiveness of RCD-based affinity propagation (AP) clustering, and demonstrate the performance of Genarris for configuration space screening of three past CSP blind test targets, shown in Figure 1. The names "Target II", "Target XIII", and "Target XXII" are unique identifiers, assigned within the first, fourth, and sixth CSP blind tests, respectively.[26,29,31]

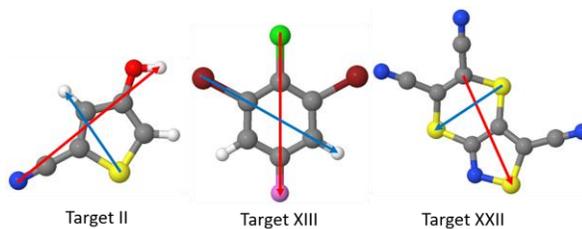

FIG. 1. Geometries of the molecules studied here. C atoms are colored in gray, H in white, N a in blue, S in yellow, O in red, Br in dark red, F in pink, and Cl in green. The red and blue arrows indicate the reference axes used to construct the relative coordinate descriptor (RCD), as described in Section II.A.1.

## II. METHODS

Genarris begins by generating crystal structures out of a single molecule 3D structure (Section II.A). The Harris approximation is used for fast screening (Section II.B). Once a large "raw" pool of structures is generated, machine learning is used for clustering based on packing motif similarity, represented by the relative positions and orientations of neighboring molecules (Section II.C). Various workflows may be used to reduce the raw pool to a small curated population by applying successive steps of energy evaluation, clustering, and selection (Section II.D).

### A. Structure Generation

#### 1. Molecule 3D Coordinates

Genarris takes as input the 3D coordinates of a single molecule. These may be generated by any means. Here, the ChemDraw software is used to obtain an estimate of the molecule's 3D atomic coordinates out of a 2D stick diagram. DFT geometry optimization is then performed using the FHI-aims electronic structure code,[94] with the Perdew-Burke-Ernzerhof (PBE)[95,96] generalized gradient approximation and the Tkatchenko-Scheffler (TS) pairwise dispersion correction.[97] *Higher-level* numerical settings are used, which correspond to the tight/tier 2 settings of FHI-aims.

## *2. Unit Cell Generation*

Unit cell generation is initialized by obtaining an estimate of the volume of a unit cell with a fixed number of molecules. 10 random structures are generated with a fixed, overestimated volume. Full unit cell relaxation is then performed using PBE+TS with *lower-level* numerical settings, which correspond to the light/tier 1 settings of FHI-aims.[94] The following parameters are used to accelerate the calculation: the k-grid is set to $2 \times 2 \times 2$, the self-consistent accuracy of eigenvalue sum is set to 0.01, and the self-consistent accuracy of forces is not checked. The smallest relaxed volume out of the set of trial structures is taken as an initial volume estimate, denoted hereafter as $u$.

Genarris uses the standard space group symmetry definitions provided by Bilbao Crystallographic Server.[98] Once the user specifies the number of molecules per cell and desired chirality (chiral or non-chiral), Genarris identifies the compatible space groups with matching general Wyckoff position multiplicity. The user may optionally specify which space group(s) to use. Additionally, special Wyckoff positions may be requested. To generate a structure, Genarris randomly picks one of the compatible or user-defined space groups.

After the space group of the random structure is determined, the lattice vectors are constructed according to the designated Bravais system. The unit cell volume may be fixed or sampled randomly within a specified range, by default between $0.9$ and $1.1u$. The user may choose to bias towards the smaller volume using a half-normal distribution curve. $0.1u$ is as the default standard deviation of the distribution. Genarris uses this design because the random placement of molecules in a unit cell with smaller volume is more difficult due to constraints imposed by close contacts.

The unit cell orientation is standardized, such that the lattice vectors, $\vec{a} = (a_x, a_y, a_z)^T$, $\vec{b} = (b_x, b_y, b_z)^T$, $\vec{c} = (c_x, c_y, c_z)^T$, form an upper triangular matrix ($a_y = a_z = b_z = 0$). The cell volume, $v$, is then given by the product of the principal components of each lattice vector: $v = a_x b_y c_z$. The user may control the cell shape by constraining the ratio between each of the principal components and the cube root of *v*. Genarris constructs the lattice vectors by randomly generating the principal components (whose product is equal to *v*) within the user-defined range. When the cell angles, α, β and γ, are not constrained by the Bravais system, Genarris randomly generates them from 30 to 150 degrees by default, or in a user defined range. Given the six parameters ($a_x, b_y, c_z, \alpha, \beta, and\ \gamma$), the unit cell is now uniquely-defined. Genarris then solves for the additional cell parameters through equations S1-S6, provided in the supplementary material.

By first ensuring reasonable principal components and then solving for the other cell parameters, Genarris effectively samples cells that are not too compressed in one direction. This leads to a higher success rate in molecule placement for skewed cell configurations, and thus increases the uniformity of sampling. This is especially important for exploring the alternative space group settings not recorded in the standard library currently implemented in Genarris. For example, the space group setting $P2_1/n$ is an alternative setting to the common space group $P2_1/c$. Expressing a structure of space group $P2_1/n$ in the $P2_1/c$ setting requires a matrix transformation of the lattice vectors, which tends to result in very oblique structures. Failure to account for this obliqueness was the reason the experimental structure of target XXII was not found with the preliminary version of our code used in the sixth blind test.[31]

### *3. Molecule Placement*

Genarris places the molecule in the asymmetric unit by giving it a random orientation and then selecting a random center of mass (COM) position. The random orientation is sampled uniformly by choosing a random rotation axis on a unit sphere (see equations S7-S10 in the supplementary material). The random rotation matrix is then applied to the molecule with its COM fixed at the origin. The COM is then moved to a random position by uniform random sampling between 0 and 1 for each dimension of the fractional coordinates. Once the asymmetric unit is constructed, the chosen space group symmetry is applied to obtain the atomic coordinates of the remaining molecules in the unit cell.

After a structure is randomly generated, a closeness check is performed to avoid unphysical close contacts. Structures that fail the closeness check are rejected. Two types of closeness checks are implemented in Genarris, a COM distance check and an intermolecular atomic distance check. The latter guarantees that no two atoms belonging to different molecules are closer than a user-defined threshold, which may be set as a constant or specific to the atomic species. The user may define a custom radius for each atom type or use the default setting of the van der Waals radii.[99] The parameter $s_r$ is a user-defined fraction of the sum of two atomic radii, such that the distance between the two atoms of different molecules cannot be smaller than $(r_1 + r_2) \times s_r$. The value of $s_r$ should be large enough to avoid unphysical structures (this is particularly important for the reliability of the HA, as discussed below) and small enough to allow for a diversity of crystal packing motifs. Genarris uses a fuzzy $s_r$ setting to increase pool diversity. $s_r$ is randomly selected at each structure generation attempt with a half-normal distribution, defined by an upper bound, standard deviation and a lower bound. The default values used here are 0.9, 0.05 and 0.8, respectively (these choices are motivated by the performance of the HA as shown in Section 4.1 below).

**B. Fast Screening with the Harris Approximation (HA)**

Within the Harris approximation,[36] the total density of a system is constructed by superposition of self-consistent fragment densities (in general, the fragments may be atoms, groups of atoms, or molecules). The DFT total energy may then be evaluated for the Harris density without performing a self-consistent cycle, providing very fast energy evaluations. This has been demonstrated as a reasonable approximation for the treatment dimers of weakly interacting molecules with dispersion-inclusive DFT in the van der Waals regime, where there is no significant density overlap or polarization.[38,100,101] Genarris uses the HA to construct the density of a molecular crystal by replicating, translating, and rotating the self-consistent density of a single molecule, which is calculated only once. This enables fast screening of initial structures using an unbiased first-principles DFT@Harris approach without resorting to force fields, which can be highly inaccurate and difficult to parametrize for atypical molecules.

To this end, we have implemented the Harris approximation in FHI-aims.[102] Others have reported similar implementations for plane-wave[38] and Gaussian[100,101] basis sets. The numeric atom-centered orbital (NAO) basis functions of FHI-aims are based on real valued linear combinations of spherical harmonics.[94] Because the spherical harmonics are fixed with respect to the *xyz*-coordinate system, rotation of a molecule produces a new linear combination of basis functions. Modified Wigner matrices[103] are employed to obtain the rotated coefficients of each basis function (a detailed account is provided in the supplementary material). The present implementation is restricted to Γ-point calculations of crystals of rigid molecules. The HA may be used in conjunction with any DFT functional and dispersion method. Here, for fast screening purposes we employ PBE+TS@Harris, where PBE is used to obtain the converged fragment densities and PBE+TS for the interactions between them. The same method was employed in the preliminary version of Genarris, used within the sixth CSP blind test.

**C. Structure Clustering**

## 1. Radial Distribution Function (RDF) and Relative Coordinate Descriptor (RCD)

Recently, there has been significant progress in formulating descriptors of molecular systems for ML purposes, such as the Coulomb matrix and the Bag of Bonds method.[62,64,104,105] Descriptors based on interatomic distances, such as pair correlation functions or distances between specific atoms are still commonly used for molecular crystals.[40,41,56,106] One such descriptor, the radial distribution function (RDF), is implemented in Genarris.[40,106] For this descriptor, the user inputs an element pair (*X*, *Y*). The RDF *G* between *X* and *Y* is defined as:

$$G_{XY}(r) = \frac{\sum_{i,j} \exp(-B(r-r_{ij})^2)}{N_X}, \qquad (1)$$

where *i* and *j* run over *X* and *Y* atoms, and $N_X$ is the number of *X* atoms. The RDF (which is a continuous function) is then sampled at a list of user-defined distance bins to form a vector descriptor. Multiple vectors of different element pairs can be concatenated to form a single RDF descriptor.

In addition to this atomic-level descriptor, we have developed the relative coordinate descriptor (RCD), intended for capturing how the molecules are positioned and oriented with respect to one another. The RCD is constructed by selecting a representative molecule and the *N* molecules with closest COM positions. *N* should be sufficiently large to correctly capture the environment of a molecule in a crystal. The default value is 16. Then, a frame of reference is constructed for each molecule. Two of the axes are vectors pointing from one fixed atom in the molecule to another (defined by user input), orthogonalized and normalized using a Gram-Schmidt procedure. The axes used here for the three targets are shown in Figure 1. The third axis is calculated as the cross product of the two user-defined axes. The relative positions are obtained by calculating the Euclidean distances between the COM positions of each of the surrounding molecules and the representative molecule and expressing them in the basis of the representative's reference frame. The relative orientations are obtained by taking the dot product between each of the three reference axes of a neighboring molecule with those of the representative molecule. The RCD of a crystal is then defined as

$$\vec{R} = \{(\vec{P^1}, \vec{Q^1}), \ldots, (\vec{P^N}, \vec{Q^N})\}, \qquad (2)$$

where $\vec{P^i}$ and $\vec{Q^i}$ are, respectively, the 3-dimensional relative position and relative orientation of the *i*[th] neighboring molecule with respect to the representative.

To compare two RCD vectors of different crystal structures, $\vec{R_1}$ and $\vec{R_2}$, an $N \times N$ matrix, ***D***, is constructed as

$$D_{i,j} = \left(\frac{|\overrightarrow{P_1^i} - \overrightarrow{P_2^j}|^2}{|\overrightarrow{P_1^i}||\overrightarrow{P_2^j}|}\right) + \frac{k}{3}(|\overrightarrow{Q_1^i} - \overrightarrow{Q_2^j}|^2), \qquad (3)$$

where *k* (by default, 1) is a parameter that enables assigning a different weight to the orientation difference and COM position difference, and 1/3 is a normalization factor. Then, the *M* smallest entries of ***D*** are selected, such that no two entries have the same *i* index or the same *j* index (For example, one may select $D_{1,3}$ and $D_{3,2}$, but not both $D_{1,3}$ and $D_{1,4}$). *M* is by default 8. The sum of the *M* entries serves as a measure of the distance between the two RCD vectors. A distance matrix is constructed for a given pool by calculating the RCD difference for all pairs of structures in the pool, using the above procedure.

## *2. Affinity Propagation Clustering*

In an initial screening workflow, clustering is useful for classifying an existing sample. For example, in the conformation-family Monte Carlo method,[55] clustering is used to monitor the overall convergence of the search. For our initial screening workflows, clustering helps maintain diversity during the selection process (see section 2.4). Genarris uses the affinity propagation (AP) clustering algorithm. While the more widely used *k*-means clustering calculates coordinate averages as cluster centers,[107] AP clustering identifies a refined set of exemplars from the initial data points.[108] This is useful for selecting representative structures from different clusters. AP clustering does not rely on a user-defined number of clusters; rather, the algorithm determines the number of clusters based on a message passing procedure between data points. The procedure is characterized by a preference value for a message to be passed from one data point to another, which can be manipulated to control the number of clusters. The result of AP clustering is consistent, in the sense that it does not depend on a randomized initialization of centers (as in *k*-means), but begins by considering all points as potential exemplars.[108] AP has also been shown to detect clusters with lower average squared distance to cluster center than *k*-centers, a version of *k*-means that similarly outputs exemplars.[108]

Genarris uses AP clustering as implemented in the scikit-learn package.[109] The input of AP clustering is a distance matrix, generated here from the RCD differences between all the structures in the pool, as explained in Section II.C.1. AP clustering outputs a cluster number for each structure, and assigns to each cluster an exemplar. By adjusting the preference value, Genarris allows the user to request either a fixed number of clusters, or the number of clusters that reaches a target silhouette score, a number between -1 to 1 that determines how well overall the structures fit into their clusters.[110] Accurate, non-overlapping clustering is characterized by a silhouette score greater than zero. A silhouette score of 0.5 or above indicates strong clustering, meaning that the algorithm identifies actual clusters, rather than arbitrarily dividing a continuous region. Once AP clustering is completed, selection procedures are available to either select the exemplars, or the structures with maximum or minimum properties within a cluster (e.g., the lowest energy), as described in Section II.D. In Section III.B it is demonstrated that AP clustering successfully identifies under-sampled clusters, a desirable behavior for the Diverse workflow of Genarris.

**D. Structure Selection Workflows**

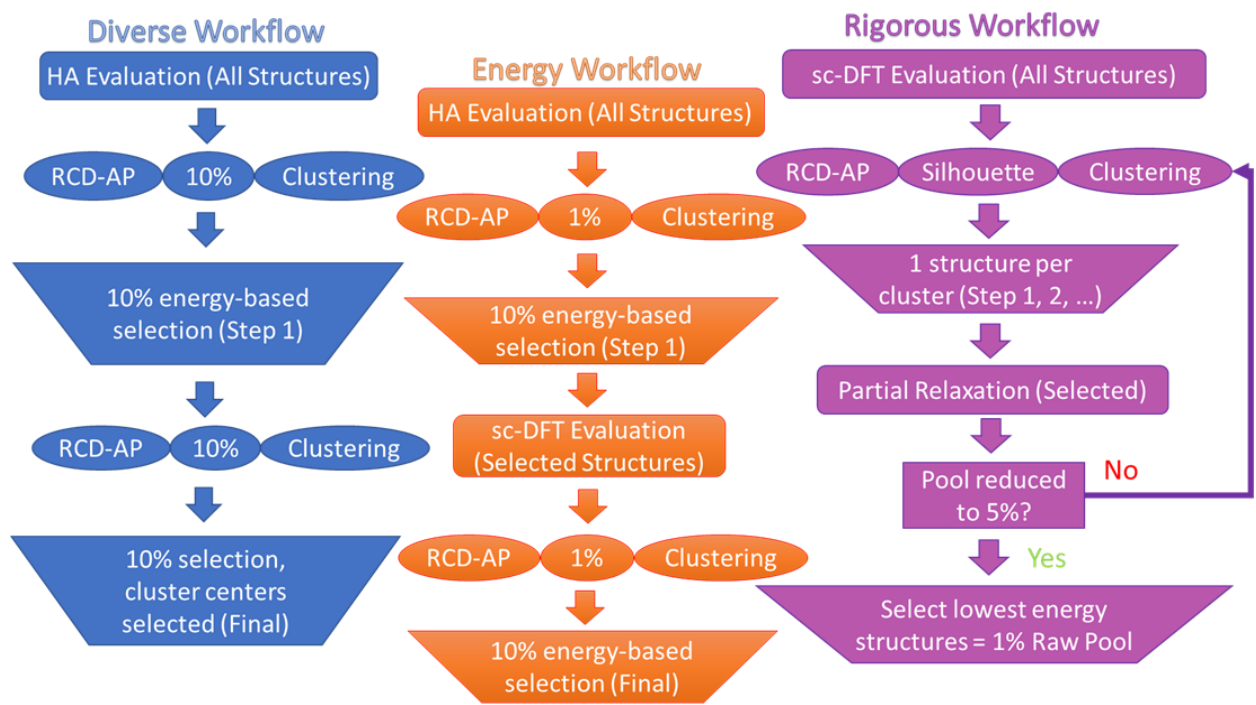

FIG. 2. Flow charts of the three screening workflows available in Genarris. RCD-AP clustering indicates AP clustering based on the RCD vector distance matrix. 1%/10% clustering means that the number of clusters is set to 1%/10% of the population. 10% energy-based selection means selecting the 10% of structures with the lowest energy within each cluster. The workflows are presented from left to right by increasing computational cost.

We have developed three standard hierarchical structure selection workflows, shown in Figure 2, whereby increasingly accurate methods are used to screen smaller subsets of structures. The workflows comprise different sequences of successive evaluation, clustering, and filtering steps. These workflows represent typical use cases of Genarris. New structure selection workflows for different purposes may be designed by the user as needed. All workflows of Genarris begin with a raw pool generated with user-defined volume range, space group symmetries, and closeness criteria, as described in Section II.A. By default, each step of the Diverse and Energy workflows reduces the pool to 10% of its previous size. All three workflows reduce the final population of structures to 1% of the raw pool. These structures may either serve directly as candidates for crystal structure prediction, or as an initial sample for a more advanced algorithm. At the end of each workflow, the final converged pool is fully relaxed, checked for duplicates, and re-ranked.

The Diverse workflow is geared towards maximally diverse sampling at a modest computational cost, intended as preparation for an advanced search algorithm. It begins by using the HA to evaluate all the structures in the raw pool. Next, RCD-based AP clustering is performed with the number of clusters set to 10% of the number of structures in the raw pool and the lowest energy structure is selected from each cluster (10% energy-based selection). This ensures the quality of the structures in the pool. Then, RCD-based AP clustering is conducted again with the number of clusters set to 10% of the remaining structures. Lastly, the exemplars chosen by the AP clustering algorithm are selected for the final pool. Because these exemplars represent the center of each cluster, they are expected to be far apart and to provide a maximally diverse sample of the configuration space.

The Energy workflow focuses on targeted sampling of low energy basins of the potential energy surface at a moderate computational cost. It creates fewer clusters than the Diverse workflow in both clustering steps in order to increase intra-cluster energy competition. Employing self-consistent DFT before the final energy-based selection improves the accuracy at the price of a higher computational cost. Like the Diverse workflow, the Energy workflow begins by using the HA to evaluate the energy of all structures in the raw pool. Next, RCD-based AP clustering is performed with the number of clusters set to 1% of the number of structures in the raw pool. The 10 lowest energy structures are selected for single point energy evaluation with FHI-aims, using PBE+TS and minimal numerical settings, where the k-grid is set to $1 \times 1 \times 1$ and the self-consistent accuracy of eigenvalue sum is set to 0.01. Then, RCD-based AP clustering is conducted with the number of clusters set to 10% of the remaining structures. Lastly, the 10 lowest energy structures in each cluster are selected for the final pool.

The Rigorous workflow is intended for exhaustive sampling of the configuration space and is essentially a standalone crystal structure prediction algorithm, based on hierarchical screening of randomly generated structures with physical constraints. It iteratively refines the pool and reduces its size. Because the Rigorous workflow fully relies on DFT for energy evaluations and structural relaxations, it requires considerable computational resources. The Rigorous workflow begins by performing single point energy evaluations for all the structures in the raw pool using PBE+TS with the *lower-level* numerical settings detailed in Section II.A.2. RCD-based AP clustering is then performed with the number of clusters adjusted to reach a silhouette score of 0.5. This value corresponds to a midpoint between barely non-overlapping clusters (silhouette score 0) and perfect clustering (silhouette score 1). Empirically, this value can consistently be reached with the number of clusters that provides a reasonable convergence rate (if a score of 0.5 cannot be reached the target score may be adjusted to a lower value). The lowest energy structure from each cluster is selected for full unit cell relaxation using PBE+TS with *lower-level* numerical settings with the number of relaxation steps constrained to 30 by default to reduce the computational cost. Through this partial relaxation, the clusters in the configuration space become more well-defined, such that the RCD-based clustering and selection process more accurately converges to a diverse and low energy post-relaxation pool. The clustering, selection, and relaxation steps are repeated until the pool size is reduced to <5% of the original sample size. At this point, we find that RCD-based clustering begins to fail as the remaining pool becomes too diverse to be reasonably clustered. Therefore, in the final step a purely energy-based selection is performed to reduce the pool size to 1% of the raw pool.

### III. COMPUTATIONAL DETAILS

Raw pools of 5,000 structures were generated for Target II and Target XIII. For Target XXII a larger pool of 10,000 structures was generated because of its conformational flexibility. The molecule can bend along the S-S axis of the six-membered ring, producing two enantiomers. The raw pools were constrained to all non-chiral space groups, with Z=4 and Z'=1. These settings correspond to the known experimental structures of the three targets. The initial volume estimates for the three targets were 546, 816, and 988 Å$^3$, respectively. The lower bound, standard deviation, and upper bound for the half-normal volume sampling (see Section II.A.2) were respectively, in units of Å$^3$, (491, 55, 600), (734, 82, 898), and (889, 99, 1098). The lower bound, standard deviation, and upper bound for the half-normal $s_r$ sampling were set to 0.80, 0.05, and 0.90 throughout. COM distance checks were conducted with minimum distances of 4, 4 and 5 Å, respectively. The RCD vectors were generated with 16 closest contacts, with reference axes selected as shown in Figure 1. For the analysis presented in Section IV.B.2, the RDF descriptor is calculated using O-N and O-S pairs, with seven 1 Å bins from 2 to 8 Å. For all workflows, the target size of the final pool was set to 1% of the raw pool size (before duplicate screening) i.e., 50, 50, and 100 structures, respectively for Targets II, XIII and XXII. The larger final pool size for Target XXII, is again because of the additional degrees of freedom associated with its conformational flexibility. The parameters used for clustering and selection are listed in Table I. For the rigorous workflow, the clustering was performed with a target silhouette score of 0.5 throughout. For the HA used in Diverse and Energy workflow, as well as in the analysis presented in Section IV.A, self-consistent single molecule calculations were performed with PBE+TS light/tier 1 settings, and crystal/dimer HA calculations were conducted with PBE+TS light/tier 1 settings, k-grid of $1 \times 1 \times 1$, and self-consistent iteration limit set to 0.

TABLE I. Clustering and selection parameters used here for the Diverse and Energy workflows.

| Workflow Step | Target II and XIII | | Target XXII | | All Targets |
|---|---|---|---|---|---|
| | No. Clusters | No. Selected Structures | No. Clusters | No. Selected Structures | No. Selected Structures per Cluster |
| Diverse Step 1 | 500 | 500 | 1000 | 1000 | 1 |
| Diverse Step 2 | 50 | 50 | 100 | 100 | 1 |
| Energy Step 1 | 100 | 500 | 200 | 1000 | 5 |
| Energy Step 2 | 10 | 50 | 20 | 100 | 5 |

For each target, the final structures produced using the Random, Diverse, and Energy workflows were used as initial pools for the GAtor genetic algorithm for molecular crystal structure prediction.[48] GAtor starts from an initial population of structures and runs several GA replicas in parallel that perform the core tasks of fitness evaluation, selection, crossover, and mutation while reading from and writing to a dynamically-updated shared population of structures. For each target, the same GA settings were used in order to compare the evolution of the different starting populations. We note that the purpose of these GA runs was not to perform an exhaustive search, for which the recommended best practice is to run GAtor several times with different settings.[48] All local optimizations within GA runs were performed with FHI-aims, using PBE+TS and *lower-level* numerical settings. For Target II, 50% standard crossover and 50% mutation were used with roulette-wheel selection and the energy-based fitness function. The GA was terminated when the common population reached at least 350 structures. For Target XIII, 50% symmetric crossover and 50% mutation were used with roulette-wheel selection and the energy-based fitness function. The GA was terminated when the common population reached at least 350 structures. For Target XXII, 50% standard crossover and 50% mutation were used with tournament selection and the energy-based fitness function. The GA was terminated when the common population reached at least 650 total structures. The GA settings used here were based on successful GA runs of these targets in Ref. 48.

## IV. RESULTS

### A. Validation of the Harris Approximation

To assess the performance of the HA for chemically diverse species with different types of intermolecular interactions, representative dimers were extracted from the experimental crystal structures of Targets II, XIII, and XXII. The intermolecular distances were varied along the closest O···N, Cl···Cl, and S···N contacts for targets II, XIII, and XXII, respectively. Figure 3 shows binding energy (BE) curves computed with self-consistent PBE+TS ($BE_{SCF}$) and PBE+TS@Harris ($BE_{HA}$), as well as the BE error, defined as: $\Delta BE(x) = BE_{HA}(x) - BE_{SCF}(x)$. The Harris density was subtracted from the self-consistent density and the residual is also shown. The HA becomes exact when the molecules are far apart and there is no interaction between them, as indicated by the asymptotic decay of the error to zero. Around the equilibrium distance, $x_{eq.}$, where the intermolecular interactions are still weak, the HA is still found to be sufficiently descriptive: The correct equilibrium distance is obtained and $|\Delta BE(x_{eq.})|$ is fairly small (0.0740, 0.0049, 0.0288 eV for targets II, XIII, and XXII, respectively). As the distance between the molecules decreases and the repulsion between their electron densities becomes significant, the assumption of non-interacting fragment densities breaks down. Because of the non-variational nature of the HA, $\Delta BE(x)$ is always negative and its magnitude increases asymptotically with decreasing distance. The decent agreement with self-consistent DFT at the equilibrium distance in the BE obtained for all three targets considered here corroborates that

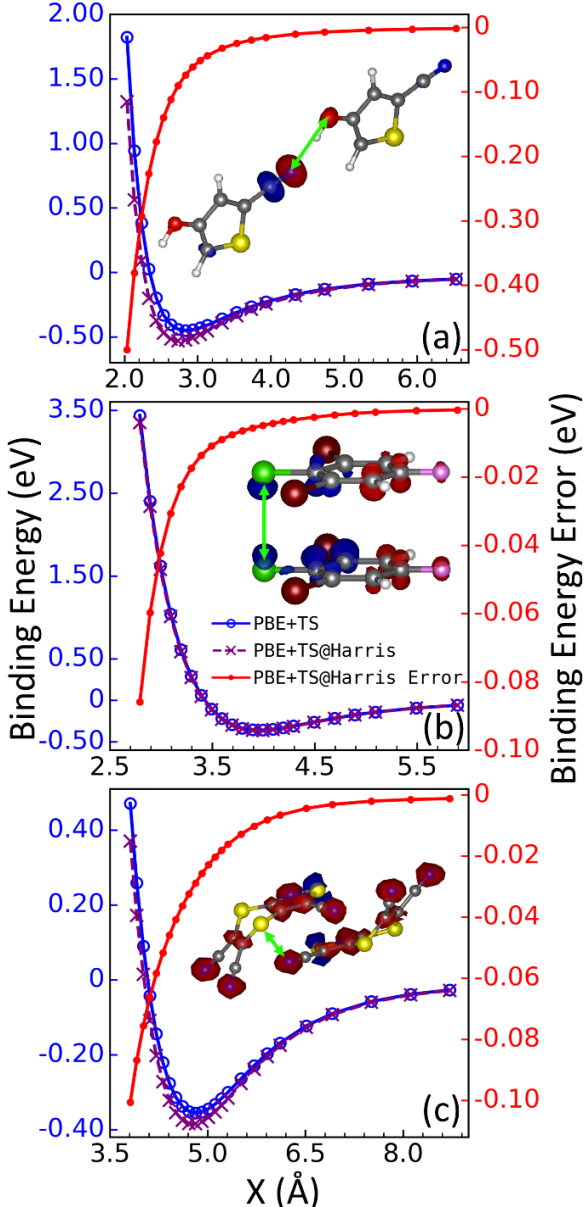

FIG. 3. Binding energy curves for dimers of (a) Target II, (b) Target XIII, and (c) Target XXII obtained using PBE+TS@Harris ($BE_{HA}$) compared to self-consistent PBE+TS ($BE_{SCF}$), and binding energy error ($BE_{HA}(x) - BE_{SCF}(x)$). The $x$ coordinate corresponds to the intermolecular O···N, Cl···Cl, and S···N distances, indicated by the green arrows. The insets show the density difference between the self-consistent and Harris densities at the equilibrium distance. Red (blue) indicates a negative (positive) density difference.

the HA is sufficiently quantitatively and (more importantly) qualitatively accurate for fast screening of the initial population of structures. These findings are consistent with earlier reports.[38,100,101,111]

Figure 3 also shows the residual difference between the self-consistent density and the Harris density at the equilibrium distance, $x_{eq.}$. Red (blue) indicates that the self-consistent density is lower (higher) than the Harris density. For Target II, the density difference is concentrated on the O and N atoms of the OH⋯N close contact, showing that the density difference due to the hydrogen bond is not captured by the HA. The strength of this bond results in a somewhat larger $|\Delta BE(x_{eq.})|$. For Target XIII, the density difference is concentrated on the six-membered ring as well as the Cl and F atoms. In this case, the HA does not capture the change in the density due to the π-π interactions between the aromatic rings and the repulsion between the halogens, which lead to the formation of a dipole with the density shifting from the F side to the Cl side of the molecule. However, the shallow BE curves indicate that these interactions are actually weak in magnitude and thus only a slight $|\Delta BE(x_{eq.})|$ is observed. For Target XXII, the density residuals suggest significant intermolecular dipole-dipole and dipole-induced-dipole interactions due to the highly polarized nitrile groups and intra-ring N atoms resulting in its moderate $|\Delta BE(x_{eq.})|$.

In the following, we further assess the reliability of the HA for energy ranking of randomly generated initial structures. The HA has not been tested in this scenario before. Three initial pools of 2,000 $P2_1/n$ structures were generated for Target XXII, using different closeness criteria with $s_r$ of 0.500, 0.625, and 0.750. Figure 4 compares the performance of PBE+TS@Harris to self-consistent PBE+TS. Panel (a) shows a direct comparison of the BE per molecule and panels (b)-(d) show the ranking based on BE per molecule from low to high. Overall, PBE+TS@Harris shows remarkable agreement with self-consistent PBE+TS for both the BEs and the rankings. The $r^2$ scores for the BEs are 0.946, 0.960, and 0.994 for $s_r$=0.500, 0.625, and 0.750, respectively. This is consistent with the above observation for dimers that the accuracy of the HA improves with increasing intermolecular distance, enforced here through a larger $s_r$ value. The $r^2$ scores for the rankings are 0.976, 0.989, and 0.979 for $s_r$=0.500, 0.625, and 0.750, respectively. Optimal performance is obtained for $s_r$=0.625. For $s_r$=0.500, the performance of the HA is worse due to the presence of more structures with unphysically close intermolecular contacts in the pool. In particular, several of the outliers exhibit unphysically close N⋯N contacts, which lead to large negative errors in the HA BEs. Three examples are circled in Figure 4 (b) and shown in panel (e). These have N⋯N distances of 1.64, 1.80 and 1.56 Å. For $s_r$=0.750 the performance of the HA deteriorates because the structures in the pool are closer in energy than in the $s_r$=0.500 and $s_r$=0.625 pools, as shown in panel (a). The accuracy of the HA is insufficient to resolve small energy differences, which leads to more ranking discrepancies.

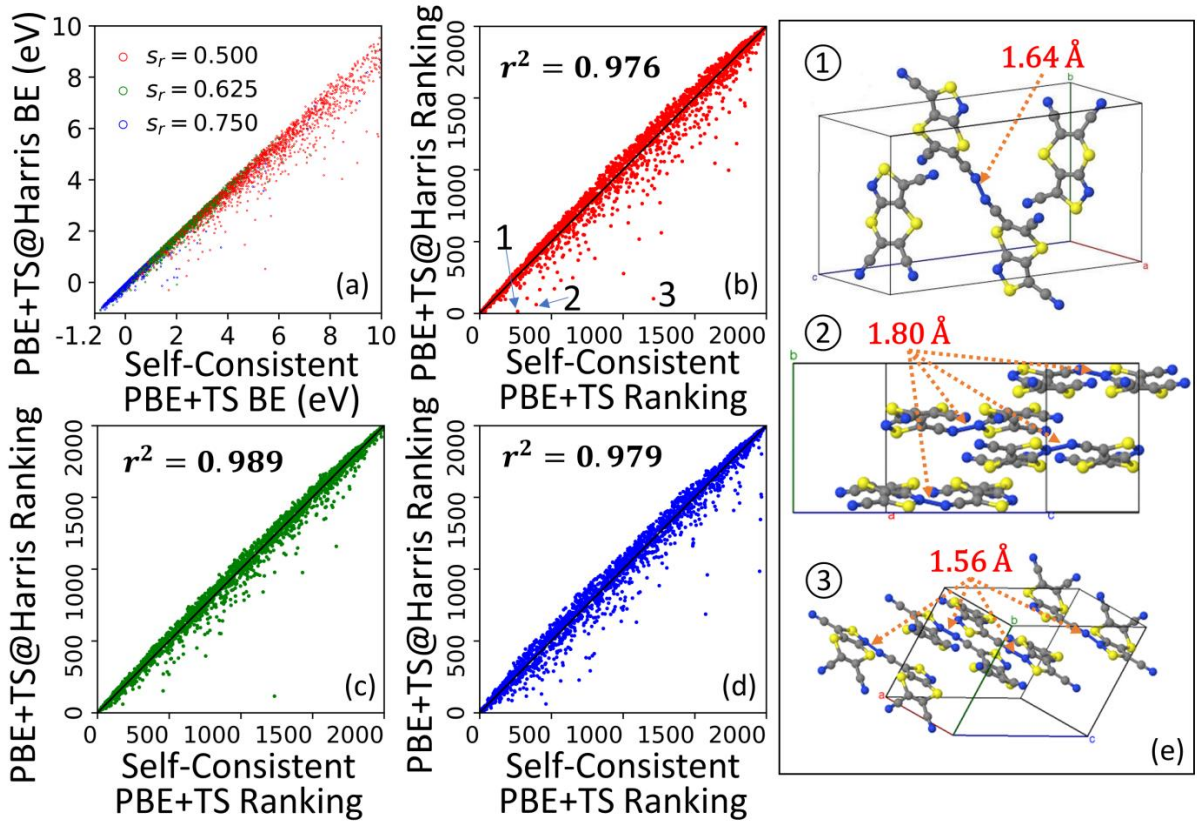

FIG. 4. (a) PBE+TS@Harris binding energy vs. self-consistent PBE+TS binding energy; (b)-(d): PBE+TS@Harris ranking vs. self-consistent PBE+TS ranking for sr=0.500, 0.625, and 0.750, respectively. The three significant outliers in (b), labeled as 1, 2 and 3, are illustrated in (e) and their unphysical N⋯N close contacts are indicated.

## B. Clustering Analysis

### 1. Comparison between k-Means and Affinity Propagation clustering

In the workflows of Genarris, AP clustering is used with respect to the RCD, as explained in Section II.C.2. Here, we illustrate the advantage of AP clustering compared to *k*-means for two dimensional and three dimensional cases, which are easier to visualize than the high dimensional RCD. To highlight the different behavior of the *k*-means and AP clustering algorithms, a set of randomly distributed points were generated within the unit circle. Construction of the data set was initiated from a few anchor points, which simulate low energy basins. Randomly generated points were then accepted or rejected based on their Euclidean distances to one of these anchor points and a random factor. Some of the anchor points had smaller random factors than others, such that fewer points were accepted in their vicinity. The resulting data set is shown in Figure 5, panel (a). The anchor points are shown as larger diamond markers. This dataset is characterized by a large, densely sampled region as well as smaller and separate satellite regions, which simulate narrow disconnected funnels of the potential energy landscape. Ideally, clustering algorithms should assign the satellite regions as distinct clusters. The results of *k*-means and AP, using 15 clusters, are shown in panels (b) and (c), respectively. While *k*-means groups three of the satellite regions together into one cluster, AP successfully identifies them as distinct clusters. This is the behavior desired by Genarris for the purpose of selecting structures from under-sampled regions of the configuration space. The high dimensional configuration space of molecular crystals often has such small clusters that are rarely explored by random sampling, for example because some packing motifs are more difficult to generate. AP clustering can correct this sampling bias by identifying these regions more effectively, provided that an appropriate descriptor is used to resolve structural differences.

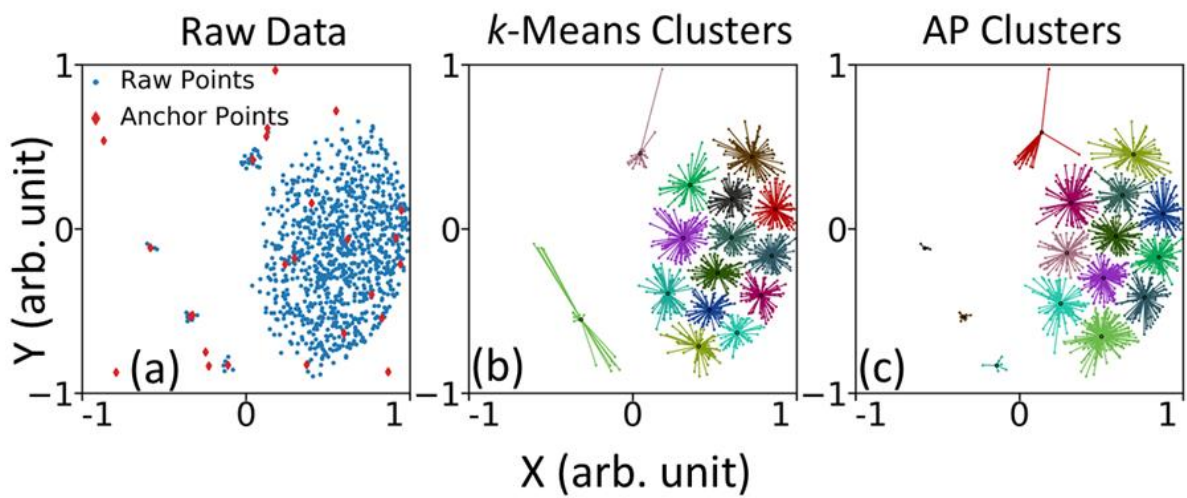

FIG. 5. Comparison of the performance of the *k*-means and AP clustering algorithms for randomly generated two-dimensional data with arbitrary units: (a) the raw data with the anchor points colored in red, and 15 clusters as found by (b) *k*-means and (c) AP.

Figure 6 compares the results of *k*-means and AP clustering algorithms in three dimensions, using a descriptor based on lattice parameters for 410 structures of Target XXII generated within the rigorous workflow. The points are grouped into five clusters by the two algorithms. Additionally, each point is colored according to the BE per molecule. The key difference between the two methods is that AP clustering identified a distinct group of structures with a low *a* parameter as a unique cluster. While the majority of the lowest energy structures are concentrated in the center of the graph, the low-*a* cluster contains structures within 0.2 eV from the respective global minimum. Therefore, it should be adequately sampled to ensure overall diversity. By identifying this region as a separate cluster, AP clustering ensures that the structures in this region are better represented in the selected pool.

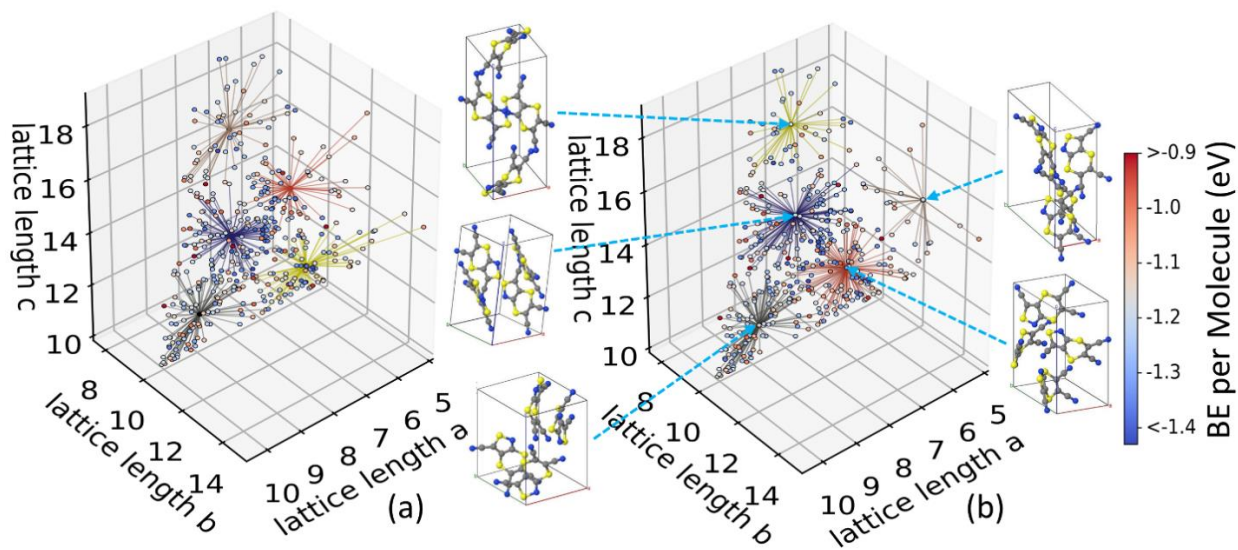

FIG. 6. Comparison of the performance of the (a) *k*-means and (b) AP clustering algorithms for 410 structures of Target XXII, clustered into five clusters with respect to a three-dimensional descriptor based on lattice parameters. Each data point is colored according to the structure's BE per molecule. The exemplars found by AP clustering are also shown.

### *2. Comparison between RDF and RCD Descriptors*

Figure 7 shows a comparison of clustering based on the RCD to clustering based on an RDF descriptor on the 5,000 random structures in the raw pool of Target II. The RCD and RDF were compared with respect to four performance measures: (a) ability to identify duplicate structures, (b) correlation with space groups, (c) correlation with unit cell volume, and (d) the silhouette score. In order to show that the differences in the clustering performance are due to the descriptor and independent of the clustering method used, both AP and *k*-means were used with the RDF descriptor (*k*-means could not be used with the RCD because its input is a conventional vector descriptor, not a distance matrix).

In the workflows of Genarris full unit cell relaxation is performed only for the final pools of structures. At this point, some structures that are similar but not identical may relax to the same structure and become duplicates. It is desirable for a descriptor to reflect the similarity between such structures, such that they are grouped into the same cluster before relaxation. For Target II, 69 pairs of duplicates were found once the final relaxed pools from the four workflows (Diverse, Energy, Rigorous, and Random) were combined. Panel (a) presents the number of duplicate pairs that were assigned to the same cluster based on their pre-relaxed geometry when the raw pool of 5,000 structures was clustered into 2-10 clusters. As a control, the raw pool was also clustered by randomly assigning a cluster number to each structure. Overall, clustering based on both descriptors significantly increases the predictive grouping of post-relaxation duplicates compared to random assignment. RCD-based clustering had a higher success rate than RDF-based clustering in assigning duplicate pairs to the same cluster. RCD-AP grouped almost all the duplicate pairs together up to 7 clusters. This helps prevent post-relaxation duplicates by eliminating them earlier in the selection process.

In panels (b)-(d) the raw pool of 5000 Target II structures was clustered into 10, 20, 40, 80, 160 and 320 clusters, based on the RCD and RDF. Panel (b) shows the number of structures whose space group is the same as the mode of its assigned cluster as a function of the number of clusters. RCD-based clustering shows a stronger correlation with space group symmetry than RDF-based clustering, which increases with the number of clusters. This indicates that RCD-based clustering captures packing motifs of molecular crystals, reflected by the space group symmetry, better than RDF-based clustering. Panel (b) shows the average intra-cluster standard deviation of unit cell volume, weighted by the number of structures in each cluster, as a function of the number of clusters. RCD-based clustering has a weaker correlation with the unit cell volume than RDF-based clustering. This trend becomes more pronounced with the number of clusters. This further demonstrates that the RCD is more sensitive to the packing motif, while the RDF is more sensitive to the unit cell volume. As previously described, the silhouette score is a measurement of how well a clustering result identifies unique clusters based on the descriptor vs. clustering a continuous region. Panel (d) shows the silhouette score as a function of the number of clusters. A higher silhouette score indicates better clustering, as explained in Section II.C.2. RCD-based clustering consistently achieves a significantly higher silhouette score than RDF-based clustering, regardless of the clustering method. Furthermore, the silhouette score for RCD-based clustering generally increases with the number of clusters, while that of RDF-based clustering decreases. This shows that the RCD provides better resolution of clusters in the configuration space. Overall, the RCD provides a superior performance to RDF, as indicated by a higher success rate in identifying duplicate structures, higher sensitivity to packing motifs, and higher silhouette scores.

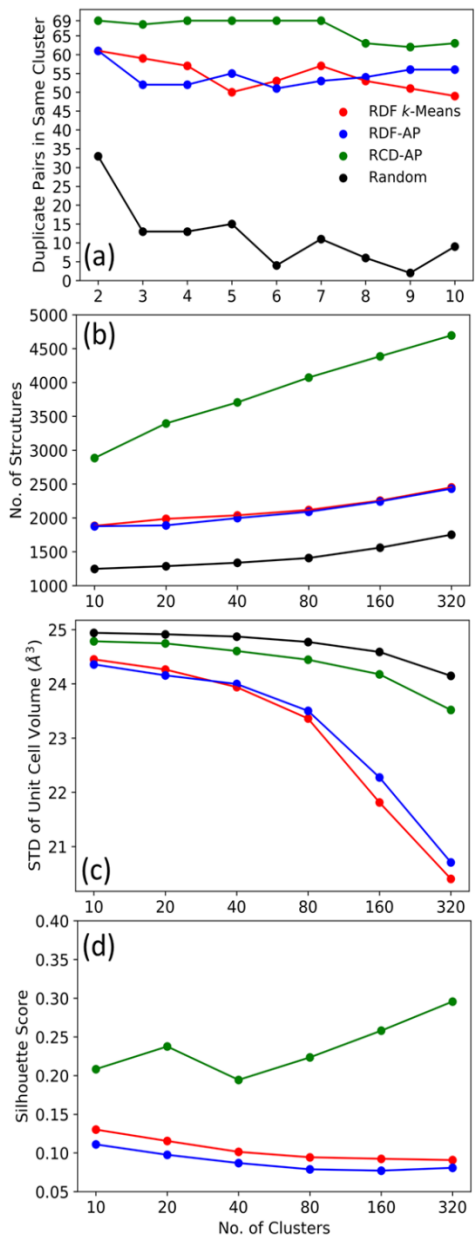

FIG. 7. Comparison of RDF-based $k$-means, RDF-based AP, and RCD-based AP clustering with respect to four metrics: (a) ability to identify duplicate structures, (b) correlation with space groups (number of structures whose space group is the same as the mode of their assigned cluster), (c) correlation with unit cell volume (intra-cluster standard deviation of unit cell volume), and (d) the silhouette score.

**C. Workflow Comparison**

Three standard workflows have been developed for Genarris, based on different sequences of successive clustering and filtering steps, as shown in Figure 2. A primary difference among the Diverse, Energy, and Rigorous workflows lies in the selection of structures from the raw pool for further evaluation and optimization. Figure 8 shows the structures selected in different steps of the three standard workflows for Targets II, XIII, and XXII. The Random workflow, used as a control, does not employ any criterion for selection. The selected structures are indicated on a graph of the PBE+TS@Harris ranking vs. the self-consistent PBE+TS ranking, plotted on a log-log scale to provide a higher resolution in the low energy region. For the Diverse and Energy workflows, the structures selected in step 1 (the first 10% selection) are highlighted in dark gray, and the final selected structures (after the second 10% selection) are highlighted in red. For the Rigorous workflow, which involves an iterative selection process, only the structures selected in the second iteration (step 2) and the final structures are highlighted in dark gray and red, respectively (additional iterations are omitted for clarity). The distributions of structures selected by the different workflows show distinct characteristics. The Energy workflow selects the majority of structures in the lower end of the spectrum for all three targets, as shown in panels (b), (f), and (j) (a few are not selected due to the clustering). Meanwhile, both the Diverse and Rigorous workflows select structure with a broader energy spectrum, with the Rigorous workflow sampling more structures in the higher energy range, as shown in panels (a), (e), (i), (c), (g), and (k). The structures sampled by the Random workflow are scattered across the distribution, with few structures ranked below 100, as shown in panels (d), (h), and (l).

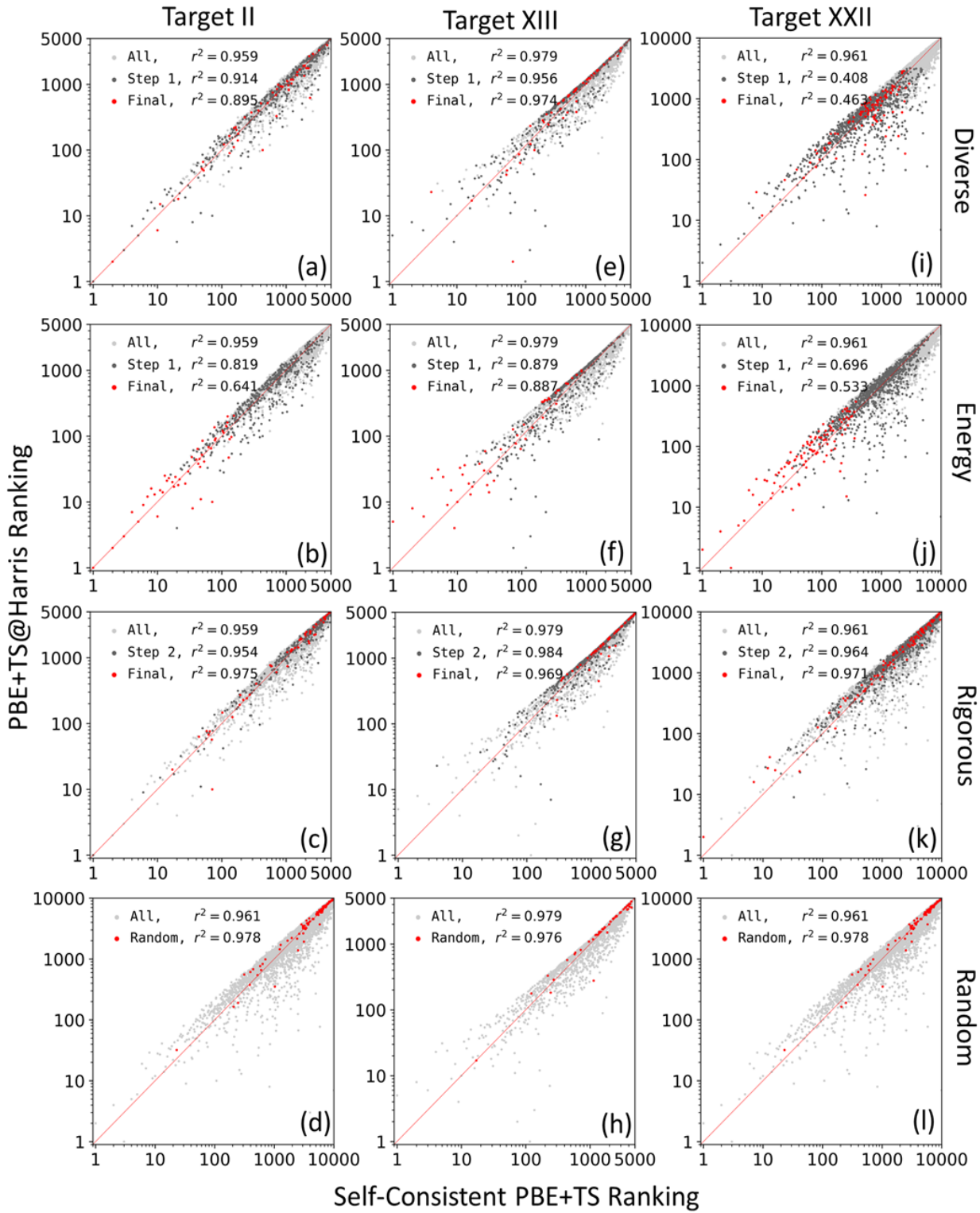

FIG. 8. PBE+TS@Harris ranking vs. self-consistent PBE+TS ranking of structures selected in the various steps of the Diverse, Energy, Rigorous, and Random workflows for Target II (panels a, b, c, d), Target XIII (panels e, f, g, h) and Target XXII (panels i, j, k, l). Selections for additional iterations of the Rigorous workflow are omitted for clarity.

The change of the $r^2$ score (calculated with self-consistent PBE+TS ranking as the "true" value and PBE+TS@Harris as "prediction") through the different workflow steps also reveals distinct patterns. In the Energy workflow $r^2$ deteriorates significantly from one step to the next because it mainly samples the lower end of the distribution, where the errors of the HA are most severe. In the Diverse workflow the deterioration of $r^2$ is typically less significant, as structures are selected across the spectrum. In the Rigorous workflow $r^2$ tends to increase in the final selection. This may be because the selection is based on self-consistent single point DFT energy evaluations, rather than on the HA. The Random workflow does not show any significant change in $r^2$, as shown in panels (d), (h), and (l). Exceptions to the $r^2$ score trends are the Energy workflow for Target XIII, shown in panel (f) and the Diverse workflow for Target XXII, shown in panel (i). In the former case, the deterioration of $r^2$ is mitigated by sampling a group of structures concentrated towards the higher end of the spectrum. The selection of structures in the higher energy region by the Energy workflow reflects the effect of clustering, which identified this region as containing distinct structural motifs that must be sampled. In the latter case, the step 1 energy-based selection of a large group of structures in the upper-middle range of the spectrum leads to a significant dip of $r^2$, which is not fully corrected by the final selection step. Additional analyses of the energy and volume distributions of the structures selected by the different workflows are provided in the supplementary material.

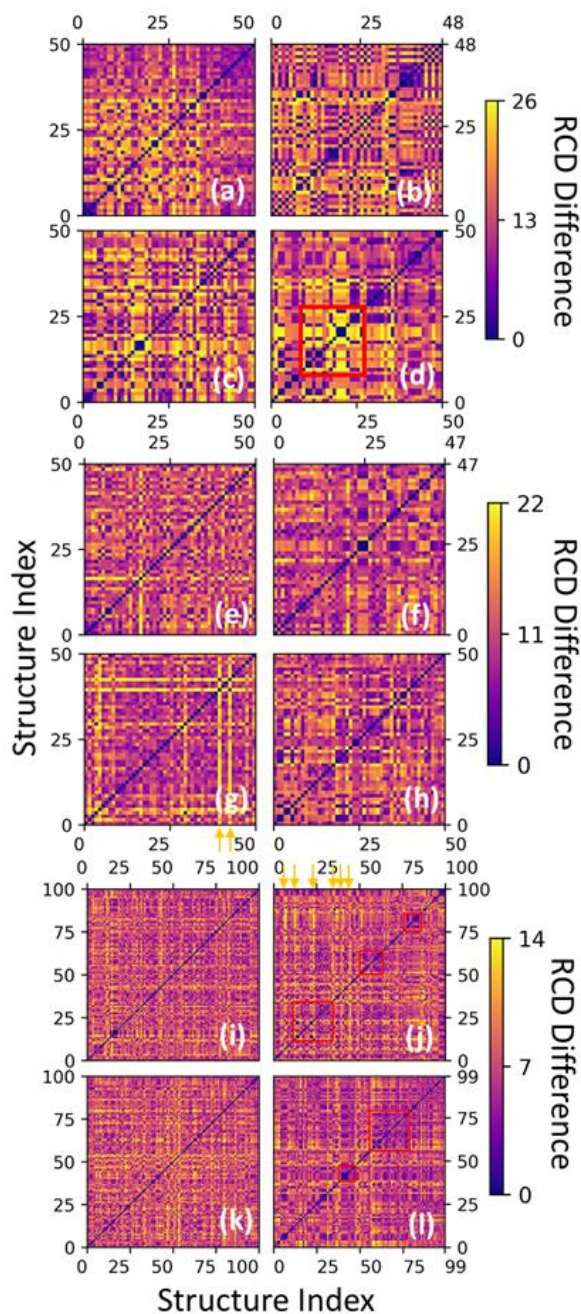

FIG. 9. Distance matrices of the Diverse, Energy, Rigorous and Random workflows for Targets II (a, b, c, d), XIII (e, f, g, h) and XXII (i, j, k l). Distances are based on the RCD as described in Section 2.3.1. The red box in (d) indicates concentrated clusters that are far from one another, and those in (j) and (l) indicate oversampled clusters. Orange arrows in (g) and (j) indicate isolated structures in under-sampled regions.

In Table II, the outcomes of the Diverse, Energy, Rigorous, and Random workflows of Genarris are compared in terms of the composition of the fully relaxed final pools of Targets II, XIII, and XXII. The Rigorous workflow successfully finds the experimental structure for all three targets, serving its purpose as a global minimum search method. The Energy workflow tends to yield a higher number of duplicates because it systematically samples the low energy regions of the potential energy surface, which increases the likelihood of sampling similar structures that relax to the same local minimum. The Diverse and Rigorous workflows tend to yield a lower number of duplicates because they are designed to sample different regions of the potential energy landscape and similar structures are effectively eliminated by clustering. Target XIII is an exception to these trends, possibly due to its halogen bonds (see also Ref. 48).

TABLE II. Analysis of the final pools for Targets II, XIII and XXII obtained with the Diverse, Energy, Rigorous and Random workflows.

| | All | Target II | | | Target XIII | | | Target XXII | | |
|---|---|---|---|---|---|---|---|---|---|---|
| Workflow | Found Exp.? | Dup. Pairs | Uniq. Struct. | Avg (STD) RCD Diff. | Dup. Pairs | Uniq. Struct. | Avg (STD) RCD Diff. | Dup. Pairs | Uniq. Struct. | Avg (STD) RCD Diff. |
| Diverse | No | 3 | 47 | 13.80 (6.46) | 2 | 48 | 11.48 (4.91) | 1 | 99 | 7.22 (2.47) |
| Energy | No | 26 | 35 | 13.82 (7.32) | 4 | 43 | 11.01 (4.56) | 28 | 80 | 7.30 (2.73) |
| Rigorous | Yes | 1 | 49 | 14.84 (7.63) | 7 | 44 | 11.76 (5.73) | 0 | 100 | 7.49 (2.53) |
| Random | No | 11 | 40 | 14.33 (7.35) | 5 | 45 | 11.11 (4.78) | 9 | 93 | 6.89 (2.55) |

The differences in the composition of the final pools produced by the Diverse, Energy, Rigorous, and Random workflows are also reflected in the distance matrices, shown in Figure 9. The structures are pre-sorted according to their BE and the distances are calculated based on the RCD, as described in Section 2.3.1. The average distance and standard deviation are given in Table II. Across the three targets, the Rigorous pools consistently have the largest average distance between structures, indicating the most diverse sampling. Graphically, this manifests as overall brighter distance matrices for Target II and XXII in panels (c) and (k). For Target XIII, the larger average may be attributed in part to the two isolated structures, appearing as two bright lines indicated by the arrows in panel (g). The distance matrices of the Energy pools have a more structured, grid-like appearance. This is particularly obvious for Targets II and XIII, as shown in panels (b) and (f). This indicates groups of structures that are similar within their clusters but different across clusters. This uneven sampling of the configuration space is reflected in the larger standard deviation of distances. For Target XXII, although the grid-like feature is not as prominent (partly due to the larger pool size), clustered sampling is revealed by the darker blocks along the diagonal, framed in red in panel (j), and isolated sampling is revealed by the bright lines, indicated by arrows. The distance matrices of the Diverse pools appear the most even and least structured, as shown in panels (a), (e), and (i). This is corroborated, especially for Targets II and XXII, by a smaller distance standard deviation, which indicates a more uniform sampling. The

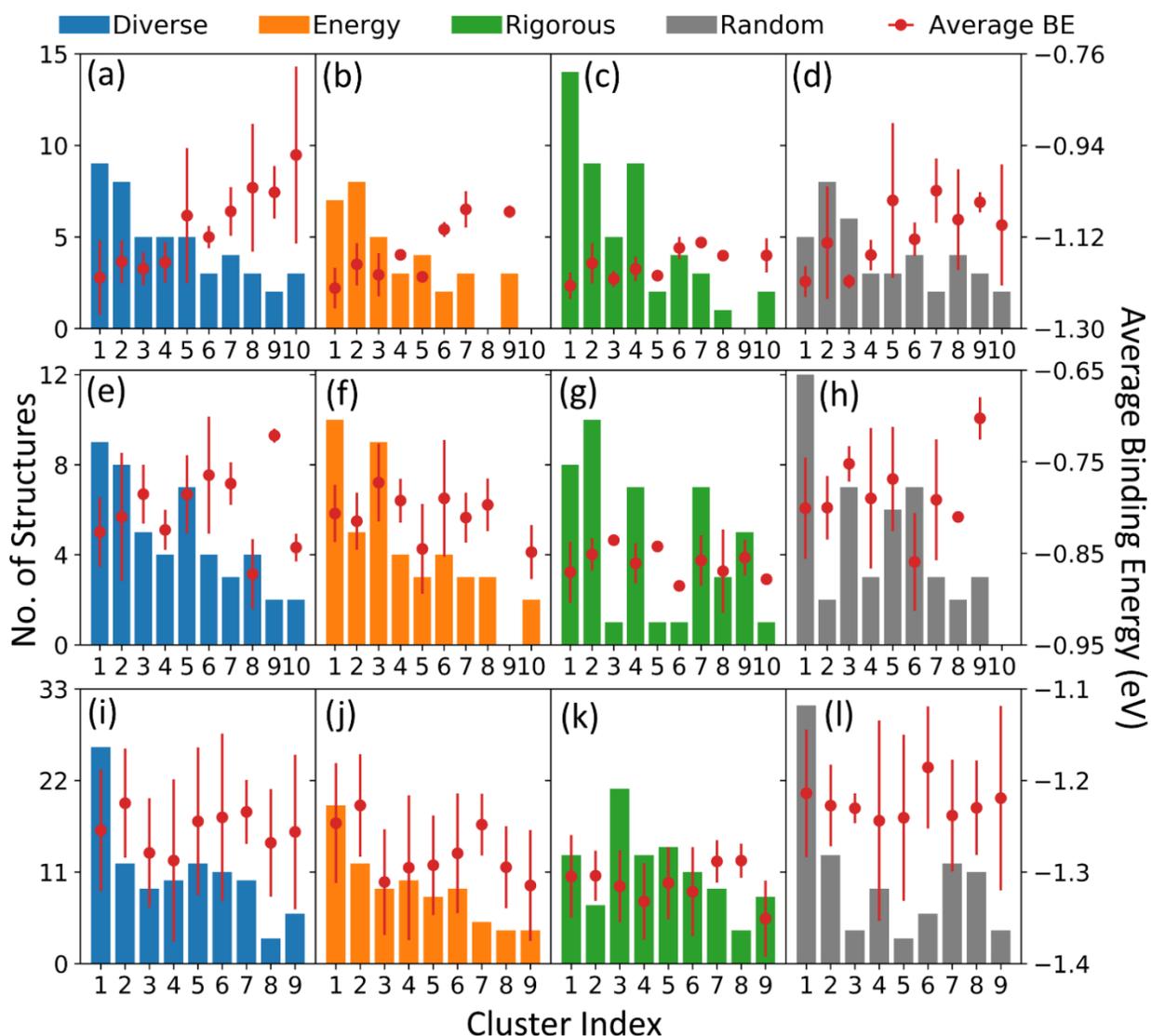

FIG. 10. Clustering analysis of the final populations generated by the Diverse, Energy, Rigorous, and Random workflow for Target II (panels a, b, c, d), Target XIII (panels e, f, g, h), and Target XXII (panels i, j, k, l). The histograms show the number of structures that fall into each cluster when the four pools are combined and clustered together. The red markers indicate the average and standard deviation of the BE per molecule for each bin.

Random pools show varied patterns in their distance matrices. For Target II, the Random workflow performed rather poorly, in terms of diverse sampling, except for the two distinct clusters in the lower energy region, framed in red in panel (d). For Targets XIII and XXII, the Random pools, shown in panels (h) and (l), exhibit similar patterns to the Energy pools, shown in panels (f) and (j). This is possibly because some basins of the configuration space are overrepresented in the raw pool and are therefore more likely to be sampled randomly.

The differences in the composition of the final pools produced by the Diverse, Energy, Rigorous, and Random workflows are further elucidated by the clustering analysis, presented in Figure 10. For this analysis, the four final workflow pools of each target were first merged, and RCD-AP clustering was applied to cluster the combined pools into 10 clusters for Target II and XIII, and 9 clusters for Target XXII. Then, histograms were generated by counting the number of structures originating from each workflow in each cluster. The average and standard deviation of the BE per molecule of the structures in each bin are also shown. Overall, the final pools of the Diverse workflow achieve the most uniform sampling across all clusters for all three targets, as shown in panels (a), (e), and (i). For Targets II and XIII, the Energy and Rigorous workflows under-sample or completely miss certain clusters, as shown in panels (b), (c), (f), and (g). The clusters under-sampled by these two energy-selective workflows tend to be higher in energy. The Rigorous workflow consistently provides the lowest energy structures with the smallest standard deviation for all three targets, as shown in panels (c), (g), and (k). In contrast, the Diverse workflow, especially for Target XXII, samples structures across a broader and higher energy range.

Overall, the results presented in this section demonstrate how the different progression of clustering and selection steps in the Diverse, Energy, and Rigorous workflows of Genarris leads to different outcomes in terms of the composition of the final pools. The selection of curated populations of structures based on different criteria may be desirable for various purposes. The user may choose one of the standard workflows suggested here or design their own workflows. Once a final population of structures is obtained, the structures may be re-relaxed and re-ranked using more accurate methods, as shown for Targets II, XIII, and XXII in Ref. 48. In addition, phonon calculations may be performed to obtain free energy ranking at finite temperatures.[112] In the next section, we demonstrate an application of Genarris for creating an initial population for a genetic algorithm and discuss the effect of the pool composition on the GA search outcomes.

**D. Genetic Algorithm Performance**

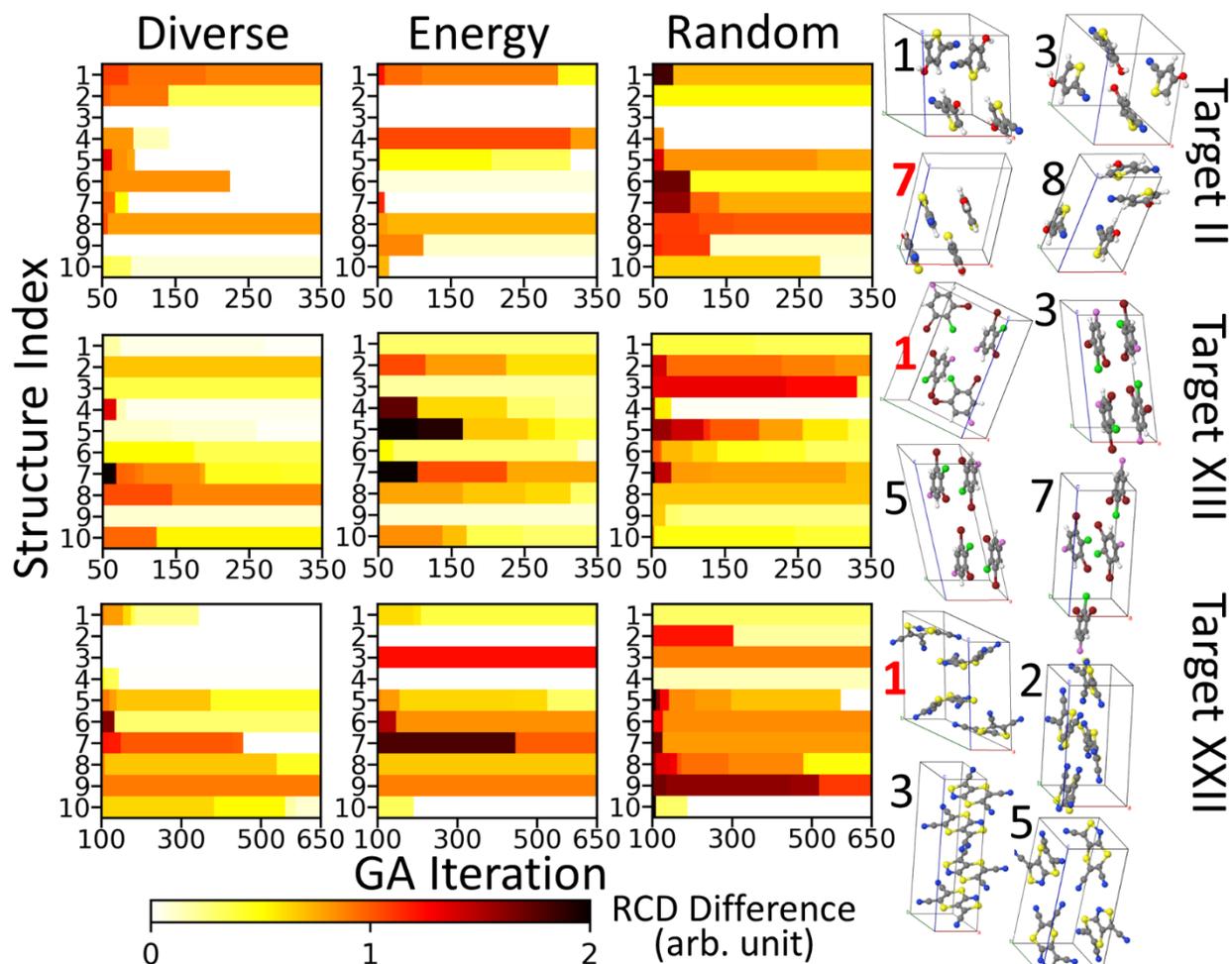

FIG. 11. Effect of the initial population on GA performance measured in terms of the minimal RCD distance to a set of representative structures as a function of GA iteration. Some of the representative structures are shown on the right with the experimental structures marked in red.

The Energy, Diverse, and Random pools generated for each target were used as initial populations for the GAtor genetic algorithm for crystal structure prediction with the settings described in section III. To illustrate the effect of the initial pool on the behavior of GAtor, a set of low energy structures, representative of the main packing motifs of each target, were selected from Ref. 48. The structures are indexed according to their relative energy, as calculated with the PBE-based hybrid functional, PBE0, and the MBD method therein. In Figure 11, the smallest RCD distance to each of these representatives is plotted as a function of GA iteration to show the convergence towards these structures for Target II (panels a, b, c), Target XIII (panels d, e, f), and Target XXII (panels h, i, j). In each panel the row colors become lighter from left to right as the GA reaches closer towards each representative structure. White indicates that the structure is found. Not all the representative structures were found by the time the GA runs sampled here were stopped (we note that the purpose of this analysis was not to perform an exhaustive GA search, as explained in Section III). The convergence towards these structures provides useful information on how the composition of the initial pool affects the GA performance.

Overall, starting the GA from the Diverse pool results in the best performance, reaching most of the representative structures and approaching the rest closely within the iteration limit used here. In particular, these runs consistently find the experimental structures (#7 for Target II and #1 for Targets XIII and XXII). Starting the GA from the Energy pool leads to inconsistent performance. The Energy pool is a good starting point for Target II, finding the experimental structure within a few iterations and also approaching most closely the #1 PBE0+MBD structure. However, for Target XIII and Target XXII, the GA runs started from Energy pool fail to reach most of the representative structures. This inconsistent performance may be a function of whether or not certain packing motifs are adequately represented in the low energy region of the raw pool, as ranked by the HA. The GA runs started from the Random pools consistently exhibit the worse performance, only reaching a few of the representative structures. We therefore conclude that starting a GA from a maximally diverse initial population provides the optimal performance. We recommend using the Diverse workflow of Genarris to produce initial pools for GAtor.

## V. CONCLUSION

We have introduced Genarris, a Python package for generating random crystal structures of rigid molecules, and demonstrated its application for three past blind test targets. For fast screening of random structures, Genarris relies on the Harris approximation (HA), which has been implemented in the FHI-aims code. The Harris density of a molecular crystal is constructed by superposition of single molecule densities, calculated only once. The DFT total energy is then evaluated for the Harris density without performing a self-consistency cycle. The HA has been validated for binding energy curves of molecular dimers as well as for the ranking of randomly generated molecular crystal structures. The HA is found to be sufficiently reliable in both scenarios, as long as no molecules are unphysically close to each other, in which case the HA fails to capture the strong repulsion between the molecular densities. This situation is avoided in Genarris by imposing a minimum distance between atoms belonging to different molecules during structure generation.

Beyond random structure generation, three standard workflows have been proposed for using Genarris to create curated populations of structures by applying successive steps of clustering and selection to the "raw" pool. The Rigorous workflow is a crystal structure prediction in and of itself, the Energy workflow creates a low-energy pool of structures, and the Diverse workflow balances low energy and maximal diversity. To perform clustering based on structural similarity within the three workflows, we have developed the relative coordinate descriptor (RCD). The RCD is based on the relative positions and orientations of neighboring molecules in the crystal, rather than on interatomic distances. Two machine learning algorithms for clustering, $k$-means and affinity propagation (AP), have been tested here, in conjunction with the RCD and a radial distribution function (RDF) descriptor. RCD-based AP clustering has been found to yield the best performance. AP clustering is better than $k$-means at resolving isolated structurally distinct clusters. RCD-based clustering is better than RDF-based clustering at identifying potential duplicates, resolving packing motif similarity (manifested as space group symmetry) rather than unit cell volume similarity, and achieving a higher silhouette score. Therefore, RCD-AP clustering is the method of choice for all workflows of Genarris.

The outcomes of the Rigorous, Energy, and Diverse workflows have been evaluated with respect to the composition of the final populations of structures and compared to a Random workflow, which selects structures randomly for the final pool. The Rigorous workflow has proven to be an effective structure search method, as it successfully located the experimentally observed structures of all three targets. Based on several indicators, the Diverse workflow provides the most uniform sampling, while the Energy workflow tends to over-sample some regions of the configuration space. The Diverse and Energy workflows have been further evaluated for the purpose of generating an initial pool of structures for a genetic algorithm. For all three targets, launching a genetic algorithm from the Diverse initial pool provides the best performance in terms of convergence towards a representative set of low-energy structures with different packing motifs.

In summary, we have demonstrated versatile applications of Genarris for random structure generation, for crystal structure prediction, and for creating an initial population of structures for a genetic algorithm. For crystal structure prediction, we suggest either using the Rigorous workflow of Genarris or using the Diverse workflow to create an initial pool, followed by the GAtor genetic algorithm. Genarris may be applied more broadly for a variety of purposes. For example, Genarris may be used to create curated sets of structures for other optimization algorithms, such as swarm algorithms, Monte Carlo methods, and Bayesian optimization, or to create training sets for machine learning algorithms. To this end, the user may choose one of the workflows proposed here or design their own workflows. In the future, Genarris will be extended to treat flexible molecules with bond-rotational degrees of freedom.

**SUPPLEMENTARY MATERIAL**

See supplementary material for additional details on the calculation of unit cell parameters and molecular rotations, additional details of the Harris approximation implementation in FHI-aims, and additional analyses of the pools of structures generated by the different workflows of Genarris.

**ACKNOWLEDGEMENTS**

Work at CMU was funded by the National Science Foundation (NSF) Division of Materials Research through grant DMR-1554428. CS, KS, and HO gratefully acknowledge support from the Solar Technologies GoHybrid initiative of the State of Bavaria. An award of computer time was provided by the Innovative and Novel Computational Impact on Theory and Experiment (INCITE) program. This research used resources of the Argonne Leadership Computing Facility, which is a DOE Office of Science User Facility supported under Contract DE-AC02-06CH11357. We thank Geoff Hutchison from the University of Pittsburgh for helpful discussions of the Harris approximation.

**REFERENCES**

[1] H. Hoppe and N.S. Sariciftci, J. Mater. Res. **19**, 1924 (2004).

[2] O. Lavastre, I. Illitchev, G. Jegou, and P.H. Dixneuf, J. Am. Chem. Soc. **124**, 5278 (2002).

[3] T. Tozawa, J.T. a Jones, S.I. Swamy, S. Jiang, D.J. Adams, S. Shakespeare, R. Clowes, D. Bradshaw, T. Hasell, S.Y. Chong, C. Tang, S. Thompson, J. Parker, A. Trewin, J. Bacsa, A.M.Z. Slawin, A. Steiner, and A.I. Cooper, Nat. Mater. **8**, 973 (2009).

[4] J.T.A. Jones, T. Hasell, X. Wu, J. Bacsa, K.E. Jelfs, M. Schmidtmann, S.Y. Chong, D.J. Adams, A. Trewin, F. Schiffman, F. Cora, B. Slater, A. Steiner, G.M. Day, and A.I. Cooper, Nature **474**, 367 (2011).

[5] J. Bernstein, Cryst. Growth Des. **11**, 632 (2011).

[6] A.J. Cruz-Cabeza, S.M. Reutzel-Edens, and J. Bernstein, Chem. Soc. Rev. **44**, 8619 (2015).

[7] J. Bernstein, *Polymorphism in Molecular Crystals* (Oxford University Press, Oxford, UK, 2010).

[8] R.K. Harris, Analyst **131**, 351 (2006).

[9] S.L. Price, D.E. Braun, and S.M. Reutzel-Edens, Chem. Commun. **52**, 7065 (2016).

[10] M. Brinkmann, G. Gadret, M. Muccini, C. Taliani, N. Masciocchi, and A. Sironi, J. Am. Chem. Soc. **122**, 5147 (2000).

[11] R.J. Tseng, R. Chan, V.C. Tung, and Y. Yang, Adv. Mater. **20**, 435 (2008).

[12] R. Pfattner, M. Mas-Torrent, I. Bilotti, A. Brillante, S. Milita, F. Liscio, F. Biscarini, T. Marszalek, J. Ulanski, A. Nosal, M. Gazicki-Lipman, M. Leufgen, G. Schmidt, W.M. Laurens, V. Laukhin, J. Veciana, and C. Rovira, Adv. Mater. **22**, 4198 (2010).

[13] M. Wang, J. Li, G. Zhao, Q. Wu, Y. Huang, W. Hu, X. Gao, H. Li, and D. Zhu, Adv. Mater. **25**, 2229 (2013).

[14] D. Yan and D.G. Evans, Mater. Horizons **1**, 46 (2014).

[15] Y. Li, D. Ji, J. Liu, Y. Yao, X. Fu, W. Zhu, C. Xu, H. Dong, J. Li, and W. Hu, Sci. Rep. **5**, 13195 (2015).

[16] C.H. Pham, E. Kucukbenli, and S. de Gironcoli, arXiv preprint arXiv:1605.00733 (2016).

[17] S.M. Woodley and R. Catlow, Nat. Mater. **7**, 937 (2008).

[18] A.R. Oganov and C.W. Glass, J. Chem. Phys. **124**, 244704 (2006).

[19] Y. Wang, J. Lv, L. Zhu, and Y. Ma, Phys. Rev. B **82**, 94116 (2010).

[20] C.J. Pickard and R.J. Needs, J. Phys. Condens. Matter **23**, 53201 (2011).

[21] C.M. Freeman, J.W. Andzelm, C.S. Ewig, J. Hill, and B. Delley, Chem. Commun. **2**, 2455 (1998).

[22] L. Stievano, F. Tielens, I. Lopes, N. Folliet, C. Gervais, D. Costa, and J.F. Lambert, Cryst. Growth Des. **10**, 3657 (2010).

[23] N. Marom, R.A. Distasio Jr., V. Atalla, S. V. Levchenko, J.R. Chelikowsky, L. Leiserowitz, and A. Tkatchenko, Angew. Chemie Int. Ed. **52**, 6629 (2013).

[24] G.J.O. Beran, Angew. Chemie Int. Ed. **54**, 396 (2015).

[25] G.J.O. Beran, Chem. Rev. **116**, 5567 (2016).

[26] J.P.M. Lommerse, W.D.S. Motherwell, H.L. Ammon, J.D. Dunitz, A. Gavezzotti, D.W.M. Hofmann, F.J.J. Leusen, W.T.M. Mooij, S.L. Price, B. Schweizer, M.U. Schmidt, B.P. van Eijck, P. Verwer, and D.E. Williams, Acta Crystallogr. Sect. B Struct. Sci. Cryst. Eng. Mater. **56**, 697 (2000).

[27] W.D.S. Motherwell, H.L. Ammon, J.D. Dunitz, A. Dzyabchenko, P. Erk, A. Gavezzotti, D.W.M. Hofmann, F.J.J. Leusen, J.P.M. Lommerse, W.T.M. Mooij, S.L. Price, H. Scheraga, B. Schweizer, M.U. Schmidt, B.P. Van Eijck, P. Verwer, and D.E. Williams, Acta Crystallogr. Sect. B **58**, 647 (2002).

[28] G.M. Day, W.D.S. Motherwell, H.L. Ammon, S.X.M. Boerrigter, R.G. Della Valle, E. Venuti, A. Dzyabchenko, J.D. Dunitz, B. Schweizer, B.P. Van Eijck, P. Erk, J.C. Facelli, V.E. Bazterra, M.B. Ferraro, D.W.M. Hofmann, F.J.J. Leusen, C. Liang, C.C. Pantelides, P.G. Karamertzanis, S.L. Price, T.C. Lewis, H. Nowell, A. Torrisi, H.A.

Scheraga, Y.A. Arnautova, M.U. Schmidt, and P. Verwer, Acta Crystallogr. Sect. B **61**, 511 (2005).

[29] G.M. Day, T.G. Cooper, A.J. Cruz-Cabeza, K.E. Hejczyk, H.L. Ammon, S.X.M. Boerrigter, J.S. Tan, R.G. Della Valle, E. Venuti, J. Jose, S.R. Gadre, G.R. Desiraju, T.S. Thakur, B.P. Van Eijck, J.C. Facelli, V.E. Bazterra, M.B. Ferraro, D.W.M. Hofmann, M.A. Neumann, F.J.J. Leusen, J. Kendrick, S.L. Price, A.J. Misquitta, P.G. Karamertzanis, G.W.A. Welch, H.A. Scheraga, Y.A. Arnautova, M.U. Schmidt, J. Van De Streek, A.K. Wolf, and B. Schweizer, Acta Crystallogr. Sect. B **65**, 107 (2009).

[30] D.A. Bardwell, C.S. Adjiman, Y. a. Arnautova, E. Bartashevich, S.X.M. Boerrigter, D.E. Braun, A.J. Cruz-Cabeza, G.M. Day, R.G. Della Valle, G.R. Desiraju, B.P. Van Eijck, J.C. Facelli, M.B. Ferraro, D. Grillo, M. Habgood, D.W.M. Hofmann, F. Hofmann, K.V.J. Jose, P.G. Karamertzanis, A. V. Kazantsev, J. Kendrick, L.N. Kuleshova, F.J.J. Leusen, A. V. Maleev, A.J. Misquitta, S. Mohamed, R.J. Needs, M. a. Neumann, D. Nikylov, A.M. Orendt, R. Pal, C.C. Pantelides, C.J. Pickard, L.S. Price, S.L. Price, H. a. Scheraga, J. Van De Streek, T.S. Thakur, S. Tiwari, E. Venuti, and I.K. Zhitkov, Acta Crystallogr. Sect. B **67**, 535 (2011).

[31] A.M. Reilly, R.I. Cooper, C.S. Adjiman, S. Bhattacharya, A.D. Boese, J.G. Brandenburg, P.J. Bygrave, R. Bylsma, J.E. Campbell, R. Car, D.H. Case, R. Chadha, J.C. Cole, K. Cosburn, H.M. Cuppen, F. Curtis, G.M. Day, R.A. DiStasio Jr, A. Dzyabchenko, B.P. van Eijck, D.M. Elking, J.A. can den Ende, J.C. Facelli, M.B. Ferraro, L. Fusti-Molnar, C.-A. Gatsiou, T.S. Gee, R. de Gelder, L.M. Ghiringhelli, H. Goto, S. Grimme, R. Guo, D.W.M. Hofmann, J. Hoja, R.K. Hylton, L. Iuzzolino, W. Jankiewicz, D.T. de Jong, J. Kendrick, N.J.J. de Klerk, H.-Y. Ko, L.N. Kuleshova, X. Li, S. Lohani, F.J.J. Leusen, A.M. Lund, J. Lv, Y. Ma, N. Marom, A.E. Masunov, P. McCabe, D.P. McMahon, H. Meekes, M.P. Metz, A.J. Misquitta, S. Mohamed, B. Monserrat, R.J. Needs, M.A. Neumann, J. Nyman, S. Obata, H. Oberhofer, A.R. Oganov, A.M. Orendt, G.I. Pagola, C.C. Pantelides, C.J. Pickard, R. Podeszwa, L.S. Price, S.L. Price, A. Pulido, M.G. Read, K. Reuter, E. Schneider, C. Schober, G.P. Shields, P. Singh, I.J. Sugden, K. Szaleqicz, C.R. Taylor, A. Tkatchenko, M.E. Tuckerman, F. Vacarro, M. Vasileiadis, A. Vazquez-Mayagoitia, L. Vogt, Y. Wang, R.E. Watson, G.A. de Wijs, J. Yang, Q. Zhu, and C.R. Groom, Acta Crystallogr. Sect. B **72**, 439 (2016).

[32] A.M. Reilly and A. Tkatchenko, Phys. Rev. Lett. **113**, 55701 (2014).

[33] F. Curtis, X. Wang, and N. Marom, Acta Crystallogr. Sect. B **72**, 562 (2016).

[34] A.M. Reilly and A. Tkatchenko, J. Chem. Phys. **139**, (2013).

[35] A. Tkatchenko, Adv. Funct. Mater. **25**, 2054 (2014).

[36] J. Harris, Phys. Rev. B **31**, 1770 (1985).

[37] G.D. Bellchambers and F.R. Manby, J. Chem. Phys. **135**, (2011).

[38] K. Berland, E. Londero, E. Schröder, and P. Hyldgaard, Phys. Rev. B **88**, 45431 (2013).

[39] D.E. Williams, Acta Crystallogr. Sect. A **52**, 326 (1996).

[40] B.P. Van Eijck and J. Kroon, J. Comput. Chem. **20**, 799 (1999).

[41] D.H. Case, J.E. Campbell, P.J. Bygrave, and G.M. Day, J. Chem. Theory Comput. **12**, 910 (2016).

[42] A. V Dzyabchenko, J. Struct. Chem. **25**, 416 (1984).

[43] I.M. Sobol, Zh. Vychisl. Mat. I Mat. Fiz. 7 **7**, 784 (1967).

[44] P.G. Karamertzanis and C.C. Pantelides, J. Comput. Chem. **26**, 304 (2005).

[45] R.G. Della Valle, E. Venuti, A. Brillante, and A. Girlande, J. Phys. Chem. A **110**, 10858 (2006).

[46] Q. Zhu, A.R. Oganov, C.W. Glass, and H.T. Stokes, Acta Crystallogr. Sect. B **68**, 215 (2012).

[47] A. Supady, V. Blum, and C. Baldauf, J. Chem. Inf. Model. **55**, 2338 (2015).

[48] F. Curtis, X. Li, T. Rose, A. Vazquez-Mayagoitia, S. Bhattacharya, L.M. Ghiringhelli, and N. Marom, to be published.

[49] R.H. Swendsen and J.-S. Wang, Phys. Rev. Lett. **57**, 2607 (1986).

[50] Y.G. Andreev, G.S. MacGlashan, and P.G. Bruce, Phys. Rev. B **55**, 12011 (1997).

[51] D.J. Earl and M.W. Deem, Phys. Chem. Chem. Phys. **7**, 3910 (2005).

[52] D.J. Wales, Science. **285**, 1368 (1999).

[53] S. Goedecker, J. Chem. Phys. **120**, 9911 (2004).

[54] G.G. Rondina and J.L.F. Da Silva, J. Chem. Inf. Model. **53**, 2282 (2013).

[55] J. Pillardy, Y.A. Arnautova, C. Czaplewski, K.D. Gibson, and H.A. Scheraga, Proc. Natl. Acad. Sci. U. S. A. **98**, 12351 (2001).

[56] J.A. Chisholm and S. Motherwell, J. Appl. Crystallogr. **38**, 228 (2005).

[57] T. Mueller, A.G. Kusne, and R. Ramprasad, Rev. Comput. Chem. **29**, 186 (2016).

[58] M. Rupp, Int. J. Quantum Chem. **115**, 1058 (2015).

[59] S. Curtarolo, D. Morgan, K. Persson, J. Rodgers, and G. Ceder, Phys. Rev. Lett. **91**, 135503 (2003).

[60] G. Ceder, D. Morgan, C. Fischer, K. Tibbetts, and S. Curtarolo, MRS Bull. **31**, 981 (2006).

[61] C.C. Fischer, K.J. Tibbetts, D. Morgan, and G. Ceder, Nat. Mater. **5**, 641 (2006).

[62] M. Rupp, A. Tkatchenko, K.-R. Müller, V. Lilienfeld, and O. Anatole, Phys. Rev. Lett. **108**, 58301 (2012).

[63] G. Montavon, M. Rupp, V. Gobre, A. Vazquez-Mayagoitia, K. Hansen, A. Tkatchenko, K.R. Müller, and O. Anatole Von Lilienfeld, New J. Phys. **15**, 95003 (2013).

[64] K. Hansen, G. Montavon, F. Biegler, S. Fazli, M. Rupp, M. Scheffler, O.A. Von Lilienfeld, A. Tkatchenko, and K.R. Müller, J. Chem. Theory Comput. **9**, 3404 (2013).

[65] G. Hautier, C.C. Fischer, A. Jain, T. Mueller, and G. Ceder, Chem. Mater. **22**, 3762 (2010).

[66] Y. Saad, D. Gao, T. Ngo, S. Bobbitt, J.R. Chelikowsky, and W. Andreoni, Phys. Rev. B **85**, 104104 (2012).

[67] K.T. Schütt, H. Glawe, F. Brockherde, A. Sanna, K.R. Müller, and E.K.U. Gross, Phys. Rev. B **89**, 205118 (2014).

[68] O. Isayev, D. Fourches, E.N. Muratov, C. Oses, K. Rasch, A. Tropsha, and S. Curtarolo, Chem. Mater. **27**, 735 (2015).

[69] A. Seko, H. Hayashi, K. Nakayama, A. Takahashi, and I. Tanaka, Phys. Rev. B **95**, 144110 (2017).

[70] M. de Jong, W. Chen, R. Notestine, K. Persson, G. Ceder, A. Jain, M. Asta, and A. Gamst, Sci. Rep. **6**, 34256 (2016).

[71] B.A. Calfa and J.R. Kitchin, AIChE J. **62**, 2605 (2016).

[72] E.O. Pyzer-Knapp, G.N. Simm, and A. Aspuru Guzik, Mater. Horiz. **3**, 226 (2016).

[73] G. Pilania, C. Wang, X. Jiang, S. Rajasekaran, and R. Ramprasad, Sci. Rep. **3**, 2810 (2013).

[74] V. Botu and R. Ramprasad, Int. J. Quantum Chem. **115**, 1074 (2015).

[75] A.P. Bartók, M.C. Payne, R. Kondor, and G. Csányi, Phys. Rev. Lett. **104**, 136403 (2010).

[76] J. Behler, J. Phys. Condens. Matter **26**, 183001 (2014).

[77] C.M. Handley and J. Behler, Eur. Phys. J. B **87**, 152 (2014).

[78] J.R. Boes, M.C. Groenenboom, J.A. Keith, and J.R. Kitchin, Int. J. Quantum Chem. **116**, 979 (2016).

[79] S.A. Ghasemi, A. Hofstetter, S. Saha, and S. Goedecker, Phys. Rev. B **92**, 45131 (2015).

[80] S. Hajinazar, J. Shao, and A.N. Kolmogorov, Phys. Rev. B **95**, 14114 (2017).

[81] A. Seko, A. Takahashi, and I. Tanaka, Phys. Rev. B **92**, 54113 (2015).

[82] J.C. Snyder, M. Rupp, K. Hansen, K.R. Müller, and K. Burke, Phys. Rev. Lett. **108**, 253002 (2012).

[83] Z.D. Pozun, K. Hansen, D. Sheppard, M. Rupp, K.R. Müller, and G. Henkelman, J. Chem. Phys. **136**, 174101 (2012).

[84] T. Stecher, N. Bernstein, and G. Csányi, J. Chem. Theory Comput. **10**, 4079 (2014).

[85] N.J. Browning, R. Ramakrishnan, O.A. von Lilienfeld, and U. Roethlisberger, J. Phys. Chem. Lett. **8**, 1351 (2017).

[86] B.R. Goldsmith, M. Boley, J. Vreeken, M. Scheffler, and L.M. Ghiringhelli, New J. Phys. **19**, 13031 (2017).

[87] S. De, F. Musil, T. Ingram, C. Baldauf, and M. Ceriotti, J. Cheminform. **9**, 1 (2017).

[88] M. Boley, B.R. Goldsmith, L.M. Ghiringhelli, and J. Vreeken, Data Min. Knowl. Discov. **31**, 1391 (2017).

[89] L.J. Nelson, G.L.W. Hart, F. Zhou, and V. Ozoliņš, Phys. Rev. B **87**, 35125 (2013).

[90] L.M. Ghiringhelli, J. Vybiral, S. V. Levchenko, C. Draxl, and M. Scheffler, Phys. Rev. Lett. **114**, 105503 (2015).

[91] L.M. Ghiringhelli, J. Vybiral, E. Ahmetcik, R. Ouyang, S. V. Levchenko, C. Draxl, and M. Scheffler, New J. Phys. **19**, (2017).

[92] P. V. Balachandran, J. Theiler, J.M. Rondinelli, and T. Lookman, Sci. Rep. **5**, 13285 (2015).

[93] E.L. Willighagen, R. Wehrens, P. Verwer, R. De Gelder, and L.M.C. Buydens, Acta Crystallogr. Sect. B **61**, 29 (2005).

[94] V. Blum, R. Gehrke, F. Hanke, P. Havu, V. Havu, X. Ren, K. Reuter, and M. Scheffler, Comput. Phys. Commun. **180**, 2175 (2009).

[95] J.P. Perdew, K. Burke, and M. Ernzerhof, Phys. Rev. Lett. **77**, 3865 (1996).

[96] J.P. Perdew, K. Burke, and M. Ernzerhof, Phys. Rev. Lett. **78**, 1396 (1997).

[97] A. Tkatchenko and M. Scheffler, Phys. Rev. Lett. **102**, 73005 (2009).

[98] M.I. Aroyo, J.M. Perez-Mato, C. Capillas, E. Kroumova, S. Ivantchev, G. Madariaga, A. Kirov, and H. Wondratschek, Zeitschrift Für Krist. Mater. **221**, 15 (2006).

[99] S. Batsanov, Inorg. Mater. **37**, 871 (2001).

[100] A. Sharapov and G. Hutchison, in *Abstr. Pap. Am. Chem. Soc.* (2012).

[101] G.R. Hutchison, in *Abstr. Pap. Am. Chem. Soc.* (2013).

[102] C. Schober, K. Reuter, and H. Oberhofer, J. Chem. Phys. **144**, 54103 (2016).

[103] M.A. Blanco, M. Flórez, and M. Bermejo, J. Mol. Struct. THEOCHEM **419**, 19 (1997).

[104] B. Huang and O.A. Von Lilienfeld, J. Chem. Phys. **145**, (2016).

[105] K. Hansen, F. Biegler, R. Ramakrishnan, W. Pronobis, O.A. Von Lilienfeld, K.R. Müller, and A. Tkatchenko, J. Phys. Chem. Lett. **6**, 2326 (2015).

[106] P. Verwer and F.J.J. Leusen, in *Rev. Comput. Chem.* (John Wiley & Sons, Inc., 2007), pp. 327–365.

[107] J.B. MacQueen, in *Proc. 5-Th Berkeley Symp. Math. Stat. Probab.* (1967), pp. 281–297.

[108] D. Dueck and B.J. Frey, Science. **315**, 972 (2007).

[109] F. Pedregosa, G. Varoquaux, A. Gramfort, V. Michel, B. Thirion, O. Grisel, M. Blondel, G. Louppe, P. Prettenhofer, R. Weiss, V. Dubourg, J. Vanderplas, A. Passos, D. Cournapeau, M. Brucher, M. Perrot, and É. Duchesnay, J. Mach. Learn. Res. **12**, 2825 (2012).

[110] P.J. Rousseeuw, J. Comput. Appl. Math. **20**, 53 (1987).

[111] Ref. 31; supplemental information of submission 12.

[112] J. Nyman, G. Day, CrystEngComm **17**, 28 (2015).